%% file: paper.tex
\documentclass[acmsmall,authordr,nonacm,screen]{acmart}

\setcopyright{rightsretained}
\copyrightyear{2024}
\acmYear{2024}
\acmDOI{}

\usepackage{stmaryrd}
\usepackage{amsmath}

\usepackage{amssymb}
\usepackage{xspace}
\usepackage{syntax}
\usepackage{semantic}
\usepackage{mathpartir}
\usepackage{tikz}
\usetikzlibrary{arrows, automata, positioning}
\usepackage{forest}
\usepackage{subcaption}
\usepackage[frozencache=true,cachedir=minted-cache]{minted}
\usepackage{wrapfig}
\usepackage{enumitem}
\usepackage{xifthen}
\usepackage{etoolbox,xpatch}

\usepackage{listings,lstlangcoq}
\lstdefinestyle{customcoq}{
  columns=flexible,
  mathescape=true,
  belowcaptionskip=1\baselineskip,
  breaklines=true,
  xleftmargin=\parindent,
  language=Coq,
  morekeywords={Variant, fun, Arguments, Type, cofix},
  emph={%
    SOCKAPI,ITree,data_at,data_at_
  },
  emphstyle={\bfseries\color{green!40!red!80}},
  showstringspaces=false,
  basicstyle=\small\ttfamily,
  keywordstyle=\bfseries\color{green!20!black},
  commentstyle=\itshape\color{red!40!black},
  identifierstyle=\color{violet!50!black},
  stringstyle=\color{orange},
  escapeinside={<@}{@>}
}

\newmintinline{c}{}
\newmintinline{coq}{}
\newcommand{\textmintinline}[2]{\text{\mintinline{#1}{#2}}}
\newcommand{\tcinline}[1]{\textmintinline{c}{#1}}

\newcommand{\ilc}[1]{\coqinline{#1}}

\newif\ifcomments\commentsfalse   %
\newif\ifaftersubmission \aftersubmissionfalse %
\newif\ifplentyofspace \plentyofspacefalse %

\include{macros}
\include{memory-macros}

\usepackage{fancyhdr}
\AtBeginDocument{%
    \addtolength{\footskip}{2.0\baselineskip}%
    \fancyfoot[L]{\textit{\copyright2024 Copyright held by the owner/author(s).}}%
}

\makeatletter
\AtBeginEnvironment{noerr}{\dontdofcolorbox}
\def\dontdofcolorbox{\renewcommand\fcolorbox[4][]{##4}}
\xpatchcmd{\inputminted}{\minted@fvset}{\minted@fvset\dontdofcolorbox}{}{}
\xpatchcmd{\mintinline}{\minted@fvset}{\minted@fvset\dontdofcolorbox}{}{} 
\makeatother
\newenvironment{noerr}{}

\begin{document}

\title{A Two-Phase Infinite/Finite Low-Level Memory Model}
\subtitle{Reconciling integer--pointer casts, finite space, and undef at the LLVM IR level of abstraction}

\author{Calvin Beck}
\orcid{0000-0002-3469-7219}
\affiliation{
  \institution{University of Pennsylvania}
  \city{Philadelphia}
  \state{PA}
  \country{USA}
}
\email{hobbes@seas.upenn.edu}

\author{Irene Yoon}
\orcid{https://orcid.org/0000-0003-3388-1257}
\affiliation{
  \institution{University of Pennsylvania}
  \country{USA}
}
\email{euisuny@cis.upenn.edu}

\author{Hanxi Chen}
\orcid{https://orcid.org/0009-0006-4486-7222}
\affiliation{
  \institution{University of Pennsylvania}
  \country{USA}
}
\email{hanxic@seas.upenn.edu}

\author{Yannick Zakowski}
\orcid{0000-0003-4585-6470}
\affiliation{
  \institution{Inria \& LIP (UMR CNRS/ENS Lyon/UCB Lyon1/INRIA)}
  \city{Lyon}
  \country{France}
}
\email{yannick.zakowski@inria.fr}

\author{Steve Zdancewic}
\orcid{0000-0002-3516-1512}
\affiliation{
  \institution{University of Pennsylvania}
  \country{USA}
}
\email{stevez@cis.upenn.edu}

\begin{abstract}
  This paper provides a novel approach to reconciling complex low-level memory
  model features, such as pointer--integer casts, with desired refinements that
  are needed to justify the correctness of program transformations. The idea is
  to use a ``two-phased'' memory model, one with and unbounded memory and
  corresponding unbounded integer type, and one with a finite memory; the
  connection between the two levels is made explicit by our notion of refinement
  that handles out-of-memory behaviors. This approach allows for more
  optimizations to be performed and establishes a clear boundary between the
  idealized semantics of a program and the implementation of that program on
  finite hardware. 

  To demonstrate the utility of this idea in practice, we instantiate the
  two-phase memory model in the context of Zakowski et al.'s \vir{}
  semantics~\cite{zakowski2021}, yielding infinite and finite memory models of
  LLVM IR, including low-level features like \tm{undef} and \tm{bitcast}. Both
  the infinite and finite models, which act as specifications, can provably be
  refined to executable reference interpreters. The semantics justify
  optimizations, such as dead-\tm{alloca}-elimination, that were previously
  impossible or difficult to prove correct.
\end{abstract}


\maketitle

\section{Introduction}

After 50 years the memory model for a programming language like C
should be well understood! Unfortunately, memory models for low-level
languages like C and LLVM IR are quite subtle and complex, especially
when considered in the context of optimizations and program
transformations~\cite{Ell12,leroy2008formal,cerebus,besson2019compcerts,besson2014precise,besson2015concrete,lee2018reconciling,besson2019verified,memarian2019exploring,kang2015formal,krebbers2014formal,krebbers2013aliasing,VIP}.  Why? These languages provide an abstract view of
memory to justify a wide range of ``high-level'' optimizations---often
pretending that available memory is unbounded and that allocations
yield disjoint blocks, where a pointer to one allocated block can
never be used to access adjacent blocks from different
allocations---but these languages also allow for low-level access to
the memory, yielding a high degree of control and
performance. Unfortunately, these two extremes are at odds, and it is
difficult to ensure that the semantics of low-level memory operations
preserve the invariants expected by the high-level optimizations.

This tension between low-level memory operations and high-level
optimizations is evident in pointer arithmetic operations.  Pointer
arithmetic operations allow programmers to manipulate memory addresses
as integer values, which exposes the underlying concrete memory
layout. When the concrete memory layout is exposed, the behavior of a
program can depend upon \textit{where} values are allocated in memory, which
can severely limit which optimizations can be performed.
For instance, the assignment \cinline{a[i] = 2} \emph{could} overwrite any other
value in memory if \cinline{a[i]} is out of bounds of the array \cinline{a} (the
underlying pointer arithmetic is hidden by the array index notation).  That
sounds reasonable when considering how this program would execute on a specific
machine at a low level, but it is disastrous from the perspective of an optimizing compiler!
Even if \cinline{a} is dead (i.e., never read from again), the compiler can't
remove this store because it \emph{might} alias with something that \emph{is}
live. To justify removing a (seemingly) dead store, the compiler would
\emph{also} have to prove that \cinline{i} is in bounds, but because \cinline{i}
can be the result of an arbitrary computation this can be difficult, if not
impossible.  What is a compiler (or compiler implementor) to do?

Programming languages like C and LLVM use the notion of
\textit{``undefined behavior''} (or ``UB'') to justify the correctness
of ``high-level'' optimizations without the need for complicated
reasoning (like determining whether \cinline{a[i]} is in bounds).  The
compiler \textit{assumes} that the program doesn't exhibit undefined
behavior.  For instance, by declaring that out-of-bounds accesses are UB,
the compiler only needs to determine that \cinline{a} isn't read from
again to justify deleting the dead \cinline{a[i] = 2} store---the
compiler assumes that \cinline{a[i]} is in bounds, so we don't need to
worry about it aliasing with anything except elements of \cinline{a}.
Programs given to the compiler must not exhibit UB or the
optimizations it performs won't be valid, potentially leading to
unexpected results, but, in return, the compiler is able to perform
much more aggressive optimizations.

While UB can be a very powerful tool, it can, unfortunately, be difficult to
define the semantics of a programming language and its memory model such that
situations that impact optimizations are classified as UB. This is
particularly challenging in the context of more realistic memory models, such as
those involving \textit{finite} memory, in which seemingly pure operations may
have visible side effects, like exhausting memory. For instance Lee, et
al. \cite{lee2018reconciling} note that, in the case of finite memory, there is
a side-channel that can be used to accurately guess the physical address of
other blocks, making it difficult to rule out pointer aliasing brought about by
casting arbitrary integers to pointers.
Similarly (but perhaps
counter-intuitively) in a finite memory model, it is, in general,
\textit{unsound} to remove a dead allocation operation: allocating less memory
can turn a program that always runs out of space into one that makes more
progress, thereby introducing more behaviors after optimization---to justify
removing the allocation, the compiler would have to also prove that the allocation always
succeeds! (This is the strategy taken in CompCertS~\cite{besson2019compcerts}.)

These situations may seem unimportant: Who cares if a program can
determine the physical address of a block without directly observing
it through a pointer--integer cast?  Isn't the goal of removing an
allocation to save memory, potentially allowing the program to make
more progress?  However, properly accounting for such semantics is
essential for ensuring that the compiler makes consistent
assumptions---inconsistency can lead to end-to-end miscompilation bugs
and subtle erroneous interactions between optimization
passes.\footnote{There are numerous discussions about such semantics
  problems on the LLVM IR github issue tracker
  \url{https://github.com/llvm/llvm-project/issues/}. A few examples
  particularly germane to our focus are
  \href{https://github.com/llvm/llvm-project/issues/54002}{54002},
  \href{https://github.com/llvm/llvm-project/issues/55061}{55061},
  \href{https://github.com/llvm/llvm-project/issues/52930}{52930},
  \href{https://github.com/llvm/llvm-project/issues/33896}{33896},
  and, especially,
  \href{https://github.com/llvm/llvm-project/issues/34577}{34577}:
  \textit{LLVM Memory Model needs more rigor to avoid undesired
    optimization results}, which has been open since 2017.}  It is also vital in the
context of formal verification, which aims for optimizations to be
``provably correct.'' For example, if the memory model does not rule
out the possibility that a program can determine the physical address
of some block, it can be impossible to justify an optimization that
depends on memory locations being unaliased. Ultimately, an incoherent
memory model leads to buggy software, and complicates formally
verifying optimization passes.

Pointer--integer casts are a major sticking point for memory
models~\cite{besson2019compcerts,kang2015formal}. These casts expose the bare bones of the memory layout, which
complicates alias analysis and can invalidate many optimizations, but
these casts also bring a more subtle and sinister issue into play:
cardinality. Most programming languages at this level of abstraction
have integers with \emph{finite} bitwidths; however, compilers and
programmers pretend that there is no limit to the number of pointers a
program can allocate, as doing so greatly simplifies reasoning (see
the discussion in Lee, et al. \cite{lee2018reconciling}).
This
discrepancy between finitary integers and infinitary pointers means
that one of the following design choices must be made: (1) a cast from
a pointer to an integer can fail, (2) casting from a pointer to an
integer and back does not necessarily yield the original pointer, and
thus causes unexpected aliasing, or, (3) we must admit that memory is
finite and make the pointer types finite as well. All of these options
have implications for program transformations: (1) means that
pointer--integer casts are effectful, instead of being (pure)
no-ops, which means they can't easily be removed, (2) means that any code
using a cast may result in pointers being truncated, which means it
could cause aliasing, thereby invalidating many optimizations, and,
(3) has many complicated consequences, including being unable to
remove dead allocations (as mentioned above).

\paragraph{The big(int) idea} This paper proposes a new way to
reconcile the desired features mentioned above.  In particular, we
present a semantics for low-level languages that provides an account
of pointer--integer casts, while dealing with finite memory and
justifying many desirable optimizations.  The key insight is that, for
such a language, there are really \textit{two} memory models in play:
one that assumes an unbounded amount of memory and the presence of an
``unbounded integer type'' (akin to \cinline{bigint}) for which there
can be an injection from pointers to these unbounded integers, and a
second that assumes a finite memory and in which all pointers and
integer types have finite bitwidths. The advantage of this is that
many optimizations, such as dead allocation elimination, are always
valid in the infinite memory model without any additional reasoning
(though some optimizations are valid in both). Optimizations should
primarily be performed with respect to the infinite memory model, and
then the program should be translated safely to the finite memory
model for subsequent lowering to machine code. This \textit{explicit}
translation step that converts the program from the infinite model to
the finite one is (almost) just the identity translation on syntax.
Semantically, it may introduce new out-of-memory behaviors, but
otherwise, the translated code retains a precisely specified
connection to its infinitary behavior.  The upshot is that we can
reason about the impact of a compiler's optimizations in phases, but
still have end-to-end guarantees about a program's behaviors that are
sufficient to ensure correct compilation. In contrast, existing models
for languages like LLVM IR don't separate these phases, which muddles
the semantics and makes the assumptions unclear. While our focus is on
languages that straddle the barrier between low-level memory accesses
and high-level optimizations like LLVM IR and C, these ideas should be
relevant to all higher-level languages that assume there is infinite
memory and must ultimately be compiled to run on machines with finite
memory.

Although this idea seems simple at first blush, working through all of the many
details turned out to be highly nontrivial---for instance, in the infinite
phase, one must address\footnote{Pun intended!} what it means to store an
``infinitely wide'' pointer into a data structure that is serialized into a form
stored in memory.  To handle that, we observe that the infinite model's
behaviors in such cases are always (eventually) lowered to the finite model's,
which gives us a degree of wiggle room: in the infinite memory model, reading
such an infinitely wide pointer ``atomically'' from memory can be lossless,
but---even in the infinite model---reading such a stored pointer byte-by-byte
can truncate the pointer.  So long as the finite model refines these behaviors,
it will behave ``as expected.''

In summary, this paper makes the following contributions:

\begin{itemize}
\item We explain the design rationale of this two-phase infinite/finite memory
  model in the context of related state-of-the-art memory models for languages
  that support integer--pointer casts.  See Section~\ref{sec:remembering-memory-models}.

\item We formalize the proposed memory model in the Coq theorem prover. The
  axiomatic description allows for many possible implementations, and we use
  Coq's module features to share the definitions common to the infinite and
  finite models.  Each model is parameterized by a few basic abstractions
  (\textit{symbolic bytes} and \textit{addresses}) that
  should be relevant in any low-level programming language. The basic interface
  is given by just seven byte-oriented operations, which can be used to build
  methods for working with aggregate data such as vectors or structs.  See
  Section~\ref{sec:memory-model-details}.

\item We define the relevant notions of simulation in both the infinite and
  finite models and prove that the translation from the infinite to the finite
  phase is a suitable refinement, introducing only new out-of-memory behaviors.

\item To demonstrate the suitability of these ideas for modeling ``real world''
  languages, we instantiate the memory models in the context of an existing
  formal semantics VIR, for LLVM IR, based on the work by Zakowski, et
  al.~\cite{zakowski2021}. The VIR semantics aims to be a specification for a
  large, practically applicable subset of LLVM IR, but its prior memory models
  suffered from the deficiencies with respect to optimization correctness
  mentioned above.  We choose this setting because VIR (and LLVM IR) supports a
  rich, C-like structured memory model, including integer--pointer casts; it
  also includes \cinline{undef} and \cinline{poison} values that interact
  non-trivially with the specification because they introduce nondeterminism and affect the notion of
  Undefined Behavior.  Our instantiation handles a superset of the features
  supported by previous versions of VIR, giving us confidence that our memory
  model scales to realistic features sets. See Section~\ref{sec:integrating-into-llvm}.

\item Besides the \textit{specification} of the memory models, which is intended
  to be a logical (and hence nondeterministic) characterization of the set of
  allowed behaviors, we also define \textit{executable implementations} for the
  VIR semantics (for both infinite and finite memories).  We formally prove that
  these implementations refine the corresponding specifications.  These
  executable semantics let us both test VIR against other LLVM IR
  implementations (specifically \texttt{clang}) and use Quickchick-style~\cite{denes2014quickchick} randomized
  testing to probe the behaviors of our model.

\item We further demonstrate the utility of this semantics for formal
  verification by proving the correctness of instances of dead-alloca
  elimination and dead \cinline{ptrtoint} cast elimination, which are
  representative of the reasoning needed to prove full-scale compiler
  optimizations such as register promotion and global-value numbering. We
  briefly describe these results in Section~\ref{sec:optimization-examples}.  
\end{itemize}

Although our proposal provides a piece of the puzzle for defining a ``full''
memory model for low-level languages like C or LLVM IR, we deliberately do not
consider some features in this paper, leaving them to other (future) work. In
particular, we omit concurrency altogether because there has already been much
research on concurrent memory models, especially for relaxed-memory semantics,
for such languages like C and the LLVM IR~\cite{BOSSW11,SV+11,CV17,compcerttso,promising,promising2}. Our treatment of
infinite--finite refinement should be orthogonal to those proposals, but we
expect it would take non-trivial engineering effort to combine them. We also
elide and/or simplify some details of LLVM IR that are not really relevant or
that we expect to be straight forward to implement using this model;
Section~\ref{sec:integrating-into-llvm} describes the features we do consider.

Our formal semantics specification, the VIR implementations, and the claimed
theorems are fully implemented and proved in the Coq interactive theorem
prover. However, the full development is very large,
relies heavily on Coq-specific details, and is a bit more general than what we
need here.  Thus, for the purposes of this paper, we have liberally simplified
and streamlined the presentation, in-lining, renaming, and sharing some
definitions when compared to the Coq code.\footnote{For reviewers: we have
  submitted an anonymized version of the repository for evaluation with the
  paper and, should the paper be accepted, we plan to submit a (cleaned up and
  better documented) version of the development for artifact evaluation. We will
  try to make the correspondence between the claims in this paper and the Coq
  code as transparent and accessible as we can, despite the simplifications made
  for the sake of presentation.}

  \section{Remembering Low-level Memory Models}
  \label{sec:remembering-memory-models}
  \input{comparing-memory-models}

\section{A Two-Phased Memory Model}
\label{sec:memory-model-details}
\input{memory-model-details}

\section {Integrating the Memory Model into an LLVM like language}
\label{sec:integrating-into-llvm}
\input{integrating-into-llvm}

\section{Optimizations Under the Memory Model}
\label{sec:optimization-examples}
\input{optimization-examples}

\section{Executable Reference Interpreters}
\label{sec:memory-model-executable}
\label{sec:testing}
\input{memory-model-executable}

\subsection{Executable \vir{}}
\input{testing}

\section{Discussion}

\subsection{Additional related work}

The individual phases in our two-phased memory model share a lot in
common with the existing state of the art in memory models---especially those already discussed in detail in Section~\ref{sec:remembering-memory-models}---but with
the crucial distinction that our approach recognizes that the
compilation pipeline for many programming languages involves a
phase-change from higher level programs with unbounded memory
semantics, to bounded machine code (a boundary which is awkwardly
straddled by compiler IRs like LLVM, and lower level languages like
C). Many projects have either an explicitly finite size of memory
\cite{LG20,besson2019compcerts,lee2018reconciling}, or utilize a
parameterized finite pointer type
\cite{krebbers2015c,krebbers2013aliasing}. C memory models
\cite{krebbers2015c,krebbers2013aliasing} often even have a
\tm{uintptr\_t} type as a parameter, which is part of the inspiration
for our \intptr{} extension to LLVM. These works generally consider a
single finite parameterization of their memory models, however, and do
not relate different parameterizations of the memory models. This
raises the question: how would the memory model with 32-bit
pointers relate to its 64-bit parameterization? We provide the answer
with our out-of-memory refinement relations, treating the
unbounded specification as the ground truth, and finite
parameterizations as refinements.

Our memory model is currently just a sequential one. Concurrent memory
models~\cite{BOSSW11,SV+11,CV17,compcerttso,promising,promising2,LG20,jung2017rustbelt}
are much more complex, but we believe the two-phased approach is
orthogonal and would apply to concurrent models as well. There are also other considerations for undefined behavior in memory
models, which we don't touch upon. In C, strict aliasing requirements
are important for ruling out pointer aliasing via the types of
pointers, which some memory models
\cite{krebbers2015c,krebbers2013aliasing} tackle. Languages like Rust
have complex ownership rules for pointers that eliminate pointer
aliasing at the type level, the semantics of which is tackled by the
RustBelt~\cite{jung2017rustbelt} project.

\subsection{The two-phase memory model in the context of \vir}

These improvements to the \vir{} semantics have been a substantial development
effort, expanding the codebase by over three fold in terms of
lines-of-Coq-code. We've aimed to keep things realistic while encapsulating the
many complications present in a substantial subset of LLVM. For instance,
\tm{undef} is known to be incredibly complicated to reason about \cite{LKSH+17},
and the \uvaluesn required to simulate \tm{undef} contain over 30 constructors,
making (proofs by) case analysis particularly arduous. Furthermore, \tm{undef}
interacts with the memory model and semantics in non-trivial ways, and many
changes were made throughout the development to figure out precisely
where \uvaluesn should undergo concretization and nondeterminism should be
collapsed so as to enable as many optimizations in the semantics as
possible. The nondeterminism in the specification monads has also been a
challenge to work with, as illustrated in the discussion surrounding
Theorem~\ref{thm:inf-to-finite-refinement}.

The product of this painstaking work is a parameterized semantics for
a substantial LLVM-like language with an in-depth characterization of
many intricate and interacting details like \tm{undef}, undefined
behavior, nondeterministic memory operations, and casts between
pointers and integers. We have done so in an effort to ease justifying
optimizations in a compiler, without the compiler itself having to
maintain complicated invariants in order to prove the validity of
important optimizations. Our verified two-phased compilation between
memory models provides a novel approach to handling the complexities
of low-level memory operations like casts between pointers and
integers in the presence of high-level optimizations, and demonstrates
the semantic necessity of considering finite memory when compiling
programs to finite architectures, which is applicable to many languages.

Having put in this effort, we are now in position to reap many rewards. For
instance the Helix project~\cite{zaliva2020verified} is a verified compiler for
a numerical programming language that targets \vir{}, and our interface should
provide a more accurate view of LLVM memory which will lend more credence to the
compilation pipeline for Helix. Similarly, our memory model should be amenable
to separation logics built using Iris~\cite{jung2018iris}, which have been used
in conjunction with \vir{} before \cite{yoon2023modular,yzz22}.
We believe that our richer memory model and higher-fidelity LLVM IR semantics
will be a boon for these and future projects that depend upon \vir{}.

\bibliographystyle{ACM-Reference-Format}
\bibliography{references}

\appendix

\end{document}

%% file: macros.tex
\newcommand{\nt}[1]{\ensuremath{\mathit{#1}}} 
\newcommand{\tm}[1]{\ensuremath{\mathtt{#1}}} 

\newcommand{\kw}[1]{\ensuremath{\mathtt{#1}}} 
\newcommand{\PROP}{\ensuremath{\mathbb{P}}}

\newcommand{\sep}{~\mid~}

\newcommand{\defeq}{\triangleq}

\newcommand{\tlist}[1]{\tm{list}~#1}

\newcommand{\vor}{~\mid~}

\newcommand{\vir}{\ensuremath{\mathrm{VIR}}\xspace}
\newcommand{\ir}{LLVM IR\xspace}

\newcommand{\itreen}{ITree\xspace}
\newcommand{\itreesn}{ITrees\xspace}

\newcommand{\itreef}[1]{\ensuremath{\tm{itree}~#1}}
\newcommand{\itree}[2]{\ensuremath{\itreef{#1}~#2}}

\newcommand{\isf}{\nt{i64}}

\newcommand{\void}{\emptyset}
\newcommand{\unit}{\ensuremath{\mathtt{(\:)}}}
\newcommand{\unitn}{\nt{unit}}  

\newcommand{\dvalue}{\ensuremath{\mathcal V}}
\newcommand{\intptr}{\ensuremath{\mathtt{iptr}}\xspace}

\newcommand{\uvaluen}{under-defined value\xspace}
\newcommand{\uvaluesn}{under-defined values\xspace}
\newcommand{\uvalue}{\ensuremath{\mathcal V_u}}
\newcommand{\undefv}{\tm{undef}}
\newcommand{\concretize}[1]{\ensuremath{\llbracket #1 \rrbracket_C}}

\newcommand{\dvint}[2]{\ensuremath{#1_{#2}\xspace}}

\newcommand{\dvinfi}[1]{\dvint{#1}{\mathit{ptr}}}

\newcommand{\dvaddr}[1]{\ensuremath{#1\xspace}}
\newcommand{\dvpoison}{\tm{poison}\xspace}
\newcommand{\dvnone}{\tm{none}\xspace}
\newcommand{\dvarray}[1]{\tm{[}#1\tm{]}\xspace}
\newcommand{\dtou}[1]{\uparrow#1}

\newcommand{\uvundef}[1]{\ensuremath{\tm{undef}_{#1}}\xspace}

\newcommand{\sbyteT}{\ensuremath{\mathcal{SB}}}
\newcommand{\SBYTE}{\tm{sbyte}}
\newcommand{\uvsbyte}[4]{\ensuremath{\SBYTE_{#1}(#2, #3, #4)}}

\newcommand{\memE}{\ensuremath{\mathcal{M}}}
\newcommand{\mpush}{\ensuremath{\tm{MPush}^{\unit}}}
\newcommand{\mpop}{\ensuremath{\tm{MPop}^{\unit}}}
\newcommand{\alloca}[1]{\ensuremath{\tm{Alloca}^{\dvalue}(#1)}}
\newcommand{\load}[2]{\ensuremath{\tm{Load}^{\uvalue}(#1,#2)}}
\newcommand{\store}[2]{\ensuremath{\tm{Store}^{\unit}(#1,#2)}}

\newcommand{\pickE}{\ensuremath{\mathcal{P}}}
\newcommand{\pick}[1]{\ensuremath{\tm{Pick}^{\dvalue}(#1)}}
\newcommand{\picku}[1]{\ensuremath{\tm{PickUnique}^{\dvalue}(#1)}}
\newcommand{\picknp}[1]{\ensuremath{\tm{PickNonPoison}^{\dvalue}(#1)}}

\newcommand{\ubE}{\ensuremath{\mathcal{U}}}
\newcommand{\ub}{\ensuremath{\tm{UB}^{\void}}}

\newcommand{\oomE}{\ensuremath{\mathcal{O}}}
\newcommand{\oom}{\ensuremath{\tm{O}^{\void}}}

\newcommand{\sem}[1]{\ensuremath{\llbracket #1 \rrbracket}}

\newcommand{\bitcast}[2]{\ensuremath{\tm{bitcast}_{#1}(#2)}}

\newcommand{\INF}[1]{\ensuremath{#1^{\var{inf}}}}
\newcommand{\FIN}[1]{\ensuremath{#1^{\var{fin}}}}

\newcommand{\eutttext}{\tm{eutt}\xspace}
\newcommand{\rutttext}{\tm{rutt}\xspace}
\newcommand{\orutttext}{\tm{orutt}\xspace}

\newcommand{\langrefine}[3]{\ensuremath{#3 \sqsupseteq_{#1} #2}}

\newcommand{\uvrefine}[2]{\ensuremath{#1 \in \concretize{#2}}}

\newcommand{\stateT}[2]{\ensuremath{\tm{stateT}_{#1}~#2}}

\newcommand{\envG}{\ensuremath{EnvG}}
\newcommand{\envL}{\ensuremath{EnvL}}
\newcommand{\mem}{\ensuremath{Mem}}



%% file: memory-macros.tex
\newcommand{\var}[1]{\ensuremath{\mathit{#1}}\xspace}
\newcommand{\None}{\ensuremath{\mathtt{None}}\xspace}
\newcommand{\Some}{\ensuremath{\mathtt{Some}}\xspace}
\newcommand{\SET}[1]{\ensuremath{\mathcal{P}(#1)}}

\newcommand{\MemPropTn}{\var{MemSpec}}
\newcommand{\MemPropT}[1]{\ensuremath{\MemPropTn(#1)}}
\newcommand{\MemExecn}{\var{MemExec}}
\newcommand{\MemExec}[1]{\ensuremath{\MemExecn(#1)}}
\newcommand{\MemState}{\ensuremath{\var{Conf}}}
\newcommand{\memState}{\varc}
\newcommand{\partialmap}[2]{#1 \hookrightarrow #2}
\newcommand{\Heap}{\var{Heap}}
\newcommand{\FrameStackType}{\var{FrameStack}}

\newcommand{\seed}{\var{used}}
\newcommand{\ptr}{\var{ptr}}
\newcommand{\Ptr}{\var{Ptr}}

\newcommand{\Memory}{\var{Memory}}

\newcommand{\MemStateFin}{\ensuremath{\MemState^{\var{fin}}}}
\newcommand{\erruboomn}{\var{Result}}
\newcommand{\erruboom}[1]{\ensuremath{\erruboomn(#1)}}
\newcommand{\UB}{\var{UB}}
\newcommand{\OOM}{\var{OOM}}
\newcommand{\FAIL}{\var{FAIL}}
\newcommand{\ok}[1]{\ensuremath{\mathit{ok}(#1)}}
\newcommand{\SByte}{\var{SByte}}
\newcommand{\addr}{\var{Addr}}
\newcommand{\Addr}{\addr}

\newcommand{\length}[1]{\lvert #1\rvert}

\newcommand{\IFF}{\ensuremath{\leftrightarrow}}
\newcommand{\IMPLIES}{\ensuremath{\rightarrow}}

\newcommand{\Frame}{\var{Frame}}

\newcommand{\TopFrame}[1]{#1.\var{top}}

\newcommand{\Provenance}{\var{Prov}}

\newcommand{\OR}{\ \ensuremath{\vee} \ }
\newcommand{\AND}{\ \ensuremath{\wedge} \ }

\newcommand{\optionn}{\var{option}}
\newcommand{\option}[1]{\ensuremath{\optionn(#1)}}

\newcommand{\refineIF}{\ensuremath{\gtrsim}}

\newcommand{\liftsbyte}{\ensuremath{\mathtt{lift\_sbyte}}}
\newcommand{\lift}[1]{\ensuremath{\lceil{}#1\rceil{}}}

\newcommand{\IntToZ}[1]{\ensuremath{\mathtt{int\_to\_}\mathbb{Z}(#1)}}

\newcommand{\ptrinf}{\ensuremath{\ptr^{\var{inf}}}}
\newcommand{\ptrfin}{\ensuremath{\ptr^{\var{fin}}}}

%% file: comparing-memory-models.tex
To put our work in context, this section provides an overview of some basic memory models,
focusing on which kinds of optimizations they allow, especially in the context of
pointer arithmetic, pointer--integer casts, and finite memory. We'll start by
reviewing a basic concrete memory model, compare it to a logical memory model
(which does not support low level memory operations, but supports more
optimizations), and then look at several memory models that bridge the gap
between these two extremes, namely, the Quasi-Concrete model
\cite{kang2015formal}, the Twin-Allocation model \cite{lee2018reconciling}, and
the CompCertS finite memory model \cite{besson2019compcerts}. The summary of the
comparison is given in Figure~\ref{fig:comparison}.

\begin{figure}
  {\small
\begin{tabular}{| l | l | l | l | p{0.3in} | p{1.5in} |}
  \hline
  \textbf{Model} & \textbf{PtoI} & \textbf{ItoP} & \textbf{Finite} & \textbf{Ext. Mem.} & \textbf{Optimizations} \\ \hline
  Concrete & No-Op & No-Op & Yes & No & Bad \\ \hline
  Logical Blocks & Unsupported & Unsupported & No & No & Good \\ \hline
  Quasi-Concrete & Effectful & Yes & No, awkward & No & Good when no PtoI casts, cannot remove PtoI casts if concrete memory is finite \\ \hline
  Twin-Allocation & No-Op & No-Op & Yes & Yes & Cannot remove dead allocations \\ \hline
  CompCertS & No-Op & No-Op & Only Finite & Yes & Have to prove optimizations use less memory \\ \hline
  Ours & No-Op & No-Op & 2-Stages & No & Staged between infinite + finite compilation allow more optimizations \\ \hline
\end{tabular}
}
\vspace{-2ex}
\caption{Comparison of various low-level memory models. Columns \textbf{PtoI} and \textbf{ItoP} describe pointer--to-int and int-to-pointer support, column \textbf{Finite} indicates support for finite memory, and \textbf{Ext. Mem.} describes whether the memory model requires extra memory. }
\label{fig:comparison}
\end{figure}

\subsection{Fully Concrete}
\label{subsec:fully-concrete}

One of the simplest memory models is a completely concrete one where
memory is modeled by an array of bytes. Each allocation is assigned
its own unique physical pointer which is just an integer index into
the memory array. Modeling pointer--integer casts under this memory
model is trivial as pointers really are just integer indices, making
the casts noops. Furthermore, finite memory can easily be modified by
simply restricting the size of the array.

\begin{wrapfigure}{r}{2in}
\begin{minted}[fontsize=\footnotesize]{c}
int main(int argc, char *argv[]) {
    char *a = malloc(4);
    char *b = malloc(4);
    *b = 1;
    char *c = a + f(0);
    *c = 2;
    // What can this print?
    printf("%d\n", *b);
    // optimized: printf("%d\n", 1)
    return 0
}
\end{minted}
\end{wrapfigure}

This model of memory is perfectly reasonable, and is quite similar to
how the memory in a physical computer actually works; unfortunately,
these memory models are \emph{too} concrete. The physical memory
layout is not abstract at all, making it difficult to justify
high level optimizations. For example, consider the program shown to the right.

It looks like \cinline{*b} is not modified after initialization, so
it's sensible to use \textit{store forwarding} to optimize away the extra load from \cinline{*b}, and
replace the call to \cinline{printf} with \cinline{printf("
Unfortunately, the simple concrete memory model can't justify this
optimization when \cinline{c == a + f(0) == b}, as this would mean that
the write to \cinline{*c} overwrites the value stored in \cinline{*b},
and so it should print \cinline{2} instead of \cinline{1}. If we want
to perform this optimization, we'll now have to know where
\cinline{a} and \cinline{b} can be allocated in memory, and what
the function call \cinline{f(0)} evaluates to. This is a lot of work
for the compiler to justify a simple optimization, especially
when the only reason it won't work is in the kind of degenerate case
where you use \cinline{a} to generate a pointer to \cinline{b}, which
should be out of bounds of \cinline{a}.

\paragraph{Undefined behavior} This is where undefined behavior (UB) comes into
play. Language designers may decide that certain behaviors are ``undefined,'' leaving the
language semantics unspecified in such cases.  The
presence of UB justifies more aggressive compiler optimizations. For instance,
in C the example program above is considered to have UB whenever \cinline{c} is
a pointer outside of the region of memory allocated for \cinline{a}. The
language implementation does not necessarily check for this UB; the
compiler simply assumes that pointers constructed using pointer arithmetic stay in bounds of the original allocation, and thus \cinline{c} could never alias with \cinline{b}, because \cinline{b} was allocated with a different call to \cinline{malloc}%
. Compilers only need to preserve defined behavior, and so any case where UB
occurs need not be considered when performing a program transformation. In the
example above, we can perform store forwarding to have
\cinline{printf("
if \cinline{c} aliases with \cinline{b}, which would make the store to
\cinline{*c} UB.

Unfortunately, this concrete memory model cannot justify such
optimizations---the model is \emph{too} well defined, giving a
specific behavior to the program in the degenerate cases where
out-of-bounds pointer arithmetic is used to overwrite arbitrary memory
locations. We shouldn't be able to use a pointer from one allocation
to derive an alias to a separate allocation. To address this we'd like
to keep track of which pointers are allowed to access which regions of
memory.

\subsection{CompCert: Provenance}
\label{subsec:compcert-provenance}

One way to solve the aliasing problem from the previous section is to give
pointers \textit{provenance}. The provenance of a pointer determines which block
in memory that pointer is allowed to access. This provenance can be preserved
throughout pointer arithmetic operations, so the pointer \cinline{c} should have
the same provenance as \cinline{a}, and thus \cinline{c} should only be able to
modify the block of memory associated with \cinline{a}, and cannot access the
disjoint block of memory from the separate allocation \cinline{b}.

One example of a memory model that takes provenance into account is CompCert's
\cite{leroy2008formal,leroy2016compcert}. CompCert is a formally verified C
compiler with an abstract-block-based memory model. Memory is no longer defined
as a concrete array; instead memory is a map of blocks, and each allocation
creates a block with a unique id \cinline{b} in the memory map. Pointers can
then be represented by a tuple \cinline{(b, o)}, where \cinline{b} is the block
id that serves as the provenance, and \cinline{o} is the offset into the block.

With this model, it's not possible for a pointer to be created that indexes into
another block, as pointer arithmetic modifies only the offset and block ids
never change. This is good news for the optimization in the example:
\cinline{c} will never be able to alias with \cinline{b}. Unfortunately, it's
not clear how we could handle casts between pointers and integers in this model
because there is no longer a physical address for a block! Furthermore, because
there is no physical memory layout, it is not clear how to implement finite
memory in this case (one could limit the total size of the
allocations, but without a physical layout of memory, it is difficult to take
fragmentation into account).

\subsection{Bringing Back Casting}

There have been a couple of proposals for how to handle pointer--integer
casting. The main two points of comparison are the quasi-concrete
\cite{kang2015formal} and twin allocation \cite{lee2018reconciling} memory
models.

\subsubsection{Quasi-concrete memory model}
\label{subsubsec:quasi-concrete}

The quasi-concrete memory model is an extension of CompCert's abstract-block
style of memory models. The memory is split into two parts: logical, and
concrete. The logical memory is the block/offset model described in
Section~\ref{subsec:compcert-provenance}, and, if no pointer--integer casts
occur, the quasi-concrete memory is effectively identical to this model.

To support casts, the quasi-concrete memory model glues a concrete memory on top
of the logical block model. This concrete layer represents the physical layout
of the blocks in memory. Whenever a pointer is cast to an integer, a physical
block is allocated in the concrete layer, representing where the logical block
is \emph{actually} allocated in physical memory.
Delaying the allocation of a physical block until cast time can rule
out situations where an address of a block might be
guessed\footnote{An address is ``guessed'' if we construct a pointer
  to a block without deriving it directly from the allocation. This is
  mostly done via integer--pointer casts. For instance, casting an
  arbitrary integer to a pointer could give you an alias to any region
  of memory. Aliasing is problematic for optimizations, so we want to
  avoid it.}. If the physical address of a block
has never been observed through a integer cast, then a
program should not be able to guess where that block is allocated (the
block exists only in logical memory). Of course, while an actual program
running on a real computer will allocate a physical address for every
block immediately, the delayed allocation of physical blocks allows
abstract pointers\footnote{Pointers whose physical address has never
  been observed via pointer--integer casts.} to be completely
isolated, such that physical addresses can never alias with them. Thus,
even in the presence of complex casting between pointers and integers,
more optimizations involving purely abstract pointers are justified, as
are simpler heuristics for aliasing.

The downside of the delayed allocation of physical blocks is that
pointer--integer casts have the side effect, within the semantics, of
allocating a physical block in concrete memory. We
cannot erase any pointer to integer casts, even if they're dead, which
in turn further restricts other optimization passes. For instance, an
otherwise dead block of code or function call may need to remain in
the program, because removing a cast will change the memory layout,
impacting the behavior of the program. Removing the cast may mean the
block is no longer accessible via integers cast to pointers, and may
change where other blocks are allocated in concrete memory.

Furthermore, the story for finite memory is awkward in the quasi-concrete memory
model. We can allocate as many logical blocks as we want, but, if there are a
finite number of physical addresses, a cast from a pointer to an integer can
cause an Out Of Memory (OOM) error. Ultimately, there are still all of the
problems that we have with reasoning about finite programs, but they arise only
in programs that perform pointer to integer casts.

\subsubsection{Twin Allocation}
\label{subsubsec:twin-allocation}

Twin allocation \cite{lee2018reconciling} takes a different approach to handling
pointer--integer casts, and does so while taking finite memory into
account. Twin allocation gives every pointer a physical address immediately, but
uses nondeterminism to rule out address guessing. Upon allocation, this memory
model actually reserves \emph{two} (or more) blocks instead of just one. One
block is a ``trap'', and accessing it will raise UB; the other will be used as
normal \footnote{Both blocks really do need to be allocated! If we just
  considered two executions, one where the block is allocated at \cinline{p1}
  and one where the block is allocated at \cinline{p2}, then it's plausible for
  something else to be allocated in the other slot for each of the executions,
  so you might not be able to swap \cinline{p1} for a trapped
  \cinline{p2}.}. The model tracks two executions for the program
nondeterministically, with the only difference between the executions being
which of the two blocks is real and which is the trap. Then, if address guessing
occurs, UB will be observed in one of the executions, as the guessed block will
instead be a trap block in the alternate execution---and in that case, the program
as a whole is considered to exhibit UB.

This model addresses some of the problems of the quasi-concrete memory
model. Casts between pointers and integers aren't effectful and can thus be
erased, as every block gets a physical address immediately. However, this model
introduces some additional constraints on program transformations. Most
importantly, allocations, even dead ones, cannot be removed! This issue is
fundamental to the nature of finite memory models: when performing any
allocation\footnote{Assuming the bound on memory size is not known in advance,
  which is the usual assumption. If the size \textit{is} known, then one could
  prove that some allocations will always succeed, but that poses other
  complications.} in a finite model, the program may run out of memory, and, if
it is removed, the program will behave differently---it might continue to
execute instead of running out of memory. This situation isn't very satisfying,
though, as programmers want the compiler to remove dead allocations!
Section~\ref{sec:memory-model-details} will discuss our solution to this
seemingly impossible conundrum.

Furthermore the twin-allocation model requires additional memory allocation to
ensure that there's enough nondeterminism for address guessing to be
detected. It should be possible for the extra allocation to be removed at run
time, as that should also yield a valid execution of the program (the ``trap''
blocks can only cause UB to be raised sooner, or OOM), but it's awkward that we
have to reason about programs with double (or more) of their actual memory
usage. Section 6 of \cite{memarian2019exploring} makes the observation that it
should be possible to instead reserve space for the largest allocation that the
program can possibly make, instead of duplicating every allocation, which makes
a slightly different reasoning trade-off. Using this strategy, one would have to
prove that a program never performs an allocation larger than this pre-allocated
trap block in order to guarantee that addresses are not guessed, which is an
additional burden on the compiler, or on the programmer if such large allocations would
be considered UB instead. 

\subsection{CompCertS: A Finite Symbolic Memory Model}
\label{subsec:compcerts}

CompCertS \cite{besson2019compcerts} extends the classic CompCert
memory model with symbolic values (as in \cite{besson2014precise} and
\cite{besson2015concrete}), and allows for pointers to be treated as
integers --- our memory model takes a very similar approach with
respect to the abstract bytes stored in memory as discussed in
Section~\ref{subsubsec:sbytes}.

For our purposes, the most relevant aspect of CompCertS is how it
handles finite memory. CompCertS makes the assumption that programs do
not run out of memory, and any program transformation that CompCertS
performs must be shown to either preserve, or decrease the amount of
memory allocated by the program. These are perfectly reasonable design
decisions, but this means that (1)  to ensure correct
compilation of a program, that program must be proven to not run out
of memory, (2) any program transformations must be shown to not use
additional memory, and (3) the finite memory address guessing
side-channel discussed in Section~\ref{subsubsec:twin-allocation} is
present.

The constraints introduced by (2) can be mitigated somewhat by
``pre-allocating'' some unused memory that can be utilized by future
program transformations, and it should also be possible to reclaim
memory that is no longer needed. Program transformations that
decrease memory usage should always be applicable (assuming the source
program does not run out of memory), but transformations that
increase memory usage may only apply conditionally. This approach to
handling finite memory is honest and yields strong guarantees about
the memory usage of the target program, but the restrictions on which
optimizations can be performed are not ideal---ideally, we want to let our
compiler hand-wave reasoning about memory altogether (in a
semantically consistent way).

%% file: memory-model-details.tex
\newcommand{\unitv}{\mathtt{tt}}

\newcommand{\singleton}[1]{\{#1\}}

\newcommand{\readbyte}[3]{\ensuremath{#1[#2]~\dot{=}~ #3}}
\newcommand{\noaccess}[2]{#1~\dot{\not\in}~#2}
\newcommand{\access}[2]{#1~\dot{\in}~#2}
\newcommand{\memokout}[3]{#1 \equiv_{\setminus#3} #2}

\newcommand{\updmap}[3]{#1\{#2 := #3\}}
\newcommand{\removemap}[2]{#1\setminus #2}

\newcommand{\varm}{\var{m}}
\newcommand{\varh}{\var{h}}
\newcommand{\varfs}{\var{fs}}
\newcommand{\varf}{\var{f}}
\newcommand{\varptr}{\var{p}}
\newcommand{\varptrs}{\overrightarrow{\var{p}}}
\newcommand{\varpr}{\var{pr}}
\newcommand{\varc}{\var{\sigma}}
\newcommand{\vara}{\var{a}}
\newcommand{\vari}{\var{i}}
\newcommand{\varb}{\var{b}}
\newcommand{\varbs}{\overrightarrow{\var{b}}}

\newcommand{\fmem}{\var{mem}}
\newcommand{\fheap}{\var{heap}}
\newcommand{\fstack}{\var{stack}}
\newcommand{\fused}{\var{used}}
\newcommand{\fadd}{\vara}
\newcommand{\fpr}{\varpr}

\newcommand{\gmemg}[1]{#1.\fmem}
\newcommand{\gmem}{\gmemg{\varc}}
\newcommand{\gmemo}{\gmemg{\varc_1}}
\newcommand{\gmemt}{\gmemg{\varc_2}}
\newcommand{\gheapg}[1]{#1.\fheap}
\newcommand{\gheap}{\gheapg{\varc}}
\newcommand{\gheapo}{\gheapg{\varc_1}}
\newcommand{\gheapt}{\gheapg{\varc_2}}
\newcommand{\gstackg}[1]{#1.\fstack}
\newcommand{\gstack}{\gstackg{\varc}}
\newcommand{\gstacko}{\gstackg{\varc_1}}
\newcommand{\gstackt}{\gstackg{\varc_2}}
\newcommand{\gusedg}[1]{#1.\fused}
\newcommand{\gused}{\gusedg{\varc}}
\newcommand{\gusedo}{\gusedg{\varc_1}}
\newcommand{\gusedt}{\gusedg{\varc_2}}
\newcommand{\gaddg}[1]{#1.\vara}
\newcommand{\gadd}{\gaddg{\varptr}}
\newcommand{\gprg}[1]{#1.\varpr}
\newcommand{\gpr}{\gprg{\varptr}}

\newcommand{\preservemem}{\gmemo = \gmemt}
\newcommand{\preserveheap}{\gheapo = \gheapt}
\newcommand{\preservestack}{\gstacko = \gstackt}
\newcommand{\preserveused}{\gusedo = \gusedt}

\newcommand{\updconf}[3]{\updmap{#1}{#2}{#3}}
\newcommand{\updconfseed}[2]{\updconf{#1}{\seed}{#2}}
\newcommand{\TailFrame}[1]{#1.tl}

\newcommand{\isbeh}[3]{(#1) \ni #2}

\newcommand{\optarg}[1]{\ifthenelse{\isempty{#1}}{}{~#1}}
\newcommand{\readbs}[2]{\mathtt{read}_b\optarg{#1}\optarg{#2}}
\newcommand{\writebs}[3]{\mathtt{write}_b\optarg{#1}\optarg{#2}\optarg{#3}}
\newcommand{\freshp}[1]{\mathtt{fresh}\optarg{#1}}
\newcommand{\findfreebk}[2]{\mathtt{find\_bk}\optarg{#1}\optarg{#2}}
\newcommand{\allocabspost}[4]{\mathtt{alloca\_post}\optarg{#1}\optarg{#2}\optarg{#3}\optarg{#4}}
\newcommand{\mallocbspost}[4]{\mathtt{malloc\_post}\optarg{#1}\optarg{#2}\optarg{#3}\optarg{#4}}
\newcommand{\allocabs}[1]{\mathtt{alloca}\optarg{#1}}
\newcommand{\mallocbs}[1]{\mathtt{malloc}\optarg{#1}}
\newcommand{\freebs}[2]{\mathtt{free}\optarg{#1}\optarg{#2}}
\newcommand{\pushfbs}[1]{\mathtt{pushf}\optarg{#1}}
\newcommand{\popfbs}[1]{\mathtt{popf}\optarg{#1}}

\newcommand{\readbe}[2]{\mathtt{read}_b^{\mathtt{run}}~#1~#2}
\newcommand{\writebe}[3]{\mathtt{write}_b^{\mathtt{run}}~#1~#2~#3}
\newcommand{\allocabe}[1]{\mathtt{alloca}^{\mathtt{run}}~#1}
\newcommand{\mallocbe}[1]{\mathtt{malloc}^{\mathtt{run}}~#1}
\newcommand{\freebe}[2]{\mathtt{free}^{\mathtt{run}}~#1~#2}
\newcommand{\pushfbe}[1]{\mathtt{pushf}^{\mathtt{run}}~#1}
\newcommand{\popfbe}[1]{\mathtt{popf}^{\mathtt{run}}~#1}

To address the limitations of the memory models described in
Section~\ref{sec:remembering-memory-models}, our proposal is to use two phases
of compilation to get the best of all worlds: an infinite memory model where
high-level abstract optimizations can be performed with ease, and a finite
memory model that more closely represents the finite architecture of the
compilation target. In our infinite model, both allocations and casts between
pointers and integers can be removed (if they're dead) or added at will, so the
presence of these operations doesn't block optimizations. Nearly all
optimizations should be done under the infinite semantics, as optimizations that
are valid under the finite model are also valid under the infinite model. Once
optimizations are performed, there is an explicit translation to the finite
model. That compilations step preserves the semantics of the original infinite
program, but potentially introduces points where the program can halt early
because it ran out of memory.

At a high level, the design of our two-phased memory model resembles
the concrete memory models from Section~\ref{subsec:fully-concrete}
and Section~\ref{subsec:compcerts}.  The only real difference between
our infinite and finite models is the type of the pointers and the
\intptr type that we introduce in Section~\ref{subsec:vir-values} in
order to handle pointer / integer casts appropriately. The infinite
model uses Coq's big-integer $\mathbb{Z}$ type for physical
addresses, and the finite versions use an implementation of 64-bit
integers (limiting the size of memory to the 64-bit address
space). The \intptr{} type matches the type of the physical addresses
in the respective memory model.

The semantics for our memory model is nondeterministic, allowing us
to accurately model the behavior of the program under the different
memory layouts that arise from nondeterministic allocations. This
nondeterminism can also be used to prevent address guessing in the
infinite memory model, as there will always be an execution where a
guessed block could be allocated somewhere else instead (akin to
swapping blocks in the twin-allocation model, except no pre-allocation
of these ``trap blocks'' is necessary because infinite memory means we
always have unallocated space to swap blocks to).

\subsection{Notations}

We write $\varm:\partialmap{A}{B}$ when $\varm$ is a partial map
from $A$ to $B$. We write $\varm[a] = b$ to assert that $a$ belongs to the
domain of $\varm$ and maps to $b$, $\updmap{\varm}{\vara}{\var{b}}$ for updating $\vara$ in $\varm$ with value
$\var{b}$, possibly extending the domain of $m$ in the process, and
$\removemap{\varm}{\vara}$ to remove $\vara$ from the domain of $\varm$.
When $\var{r}$ is an element of a record type, we write $\var{r}.\var{f}$
for the content of its field $\var{f}$. We use the notation $\SET{A}$ for the set of all subsets
of elements of $A$. Given a list, \var{l}, we write $\length{\var{l}}$ for
its length, $\var{l}[\vari]$ to access its \var{i}-th element, assuming it is
within bounds, and coerce it into a set implicitly when needed. Finally, we conflate equality
and extensional equality over finite maps.

\subsection{Memory Configurations}
\label{sec:configurations}

\begin{figure}
  \begin{minipage}[l]{.48\textwidth}
$
\begin{array}{{rcl}}
\varc\in\MemState{} &\defeq& \left\{ \begin{array}{{ll}}
                                       \fmem & : \Memory, \\
                                       \fheap & : \Heap, \\
                                       \fstack & : \FrameStackType, \\
                                       \fused & : \SET{\Provenance}
                                 \end{array} \right.
\end{array}
 $
  \end{minipage}
    \begin{minipage}[c]{.48\textwidth}
$
      \begin{array}{{rcl}}
        \varm\in\Memory & \defeq & \partialmap{\Addr}{(\SByte \times \Provenance)} \\
        \varh\in\Heap & \defeq & \partialmap{\Addr}{\SET{\Ptr}} \\
        \varptr\in\Ptr & \defeq & \{ \fadd : \Addr;~\fpr : \Provenance \}\\
        \varf\in\Frame & \defeq & \SET{\Ptr} \\
        \varfs\in\FrameStackType{} & \defeq & \tlist{\Frame} \\
        \varpr\in\Provenance{} & \defeq & \option{\mathbb{N}} \\ \\
    \end{array}
$
\end{minipage}
\vspace{-5ex}
  \caption{Datatype of memory configurations, where \Addr{} and \SByte{} are abstract parameters}
  \label{fig:memory-types}
  \vspace{-3ex}
\end{figure}

Figure~\ref{fig:memory-types} describes the datatype \MemState{} of
memory configurations. It is parameterized by two types: \Addr{}, the
representation of concrete addresses, and \SByte{}, the representation
of (symbolic) bytes, in memory. The former is straightforward:
addresses are represented as unbounded integers at the infinite level,
and bounded integers at the finite level---we assume an operation
$+:\Addr{}\rightarrow\mathbb{N}\rightarrow \OOM + \Addr{}$, which
performs arithmetic on addresses returning $\OOM$ in the case where an
overflow occurs in the finite model. We will leave the
representation of \emph{symbolic} bytes abstract, as their
implementation is language dependent, but we will give a full
description of them in our LLVM semantics in
Section~\ref{subsubsec:sbytes}; we invite the reader to think of
them as concrete bytes until then.

A configuration \memState{} has four fields. The memory itself,
$\gmem$, is a finite map from addresses to bytes with an
associated provenance. The provenance is an optional natural number, where the
\None{} constructor is used as a wildcard during integer--pointer casts.
We introduce notations to access the memory via pointers, i.e., addresses tagged
with provenance information.
We write  $\readbyte{m}{\varptr}{\varb}$ for the partial allowed $\SByte$
lookup operation:
it asserts both that $\gadd$ is in the domain of $\varm$, \emph{and} performs a
provenance check by ensuring that it maps to the pair $(\varb,\gpr)$.
We simply write $\access{\varptr}{\varm}$ as a shortcut to $\exists \varb,
\readbyte{\varm}{\varptr}{\varb}$ to state that a pointer is accessible in memory.
Conversely, $\noaccess{\varptr}{\varm}$ states that a pointer cannot be accessed
in memory, either because $\gadd$ is not in the domain of $\varm$, or because the
associated provenance is different from $\gpr$. Finally, we write
$\memokout{\varm_1}{\varm_2}{\varptrs}$ to express that memories $\varm_1$ and
$\varm_2$ agree on content and provenance at all addresses except those in the list
of pointers $\varptrs$:
\[
  \memokout{\varm_1}{\varm_2}{\varptrs}\defeq\forall \varptr', \varb, (\forall
  \varptr \in \varptrs,~\gaddg{\varptr'}\not = \gaddg{\varptr}) \IMPLIES (\readbyte{\varm_1}{\varptr'}{\varb} \IFF \readbyte{\varm_2}{\varptr'}{\varb})
\]

The heap $\gheap$ tracks information about heap allocation units.
Via \emph{malloc}, a contiguous region of memory can be allocated in
\textit{blocks} of sequentially consecutive
pointers: $\varptr_1, \ldots , \varptr_n$.  Each $\varptr_i : \Ptr$ in the block is
associated with its \textit{root} address $\gaddg{\varptr_1}$, which is the address
returned by the allocation operation. Freeing the root deallocates the whole
block. (Freeing a non-root address will be undefined behavior.)
The stack, $\gstack$, keeps track of the call stack by maintaining
a stack of \emph{frames}, where each frame consists of a list of pointers.
Fresh addresses, allocated via \tm{alloca}, are added to the top frame of the
stack, referred to as $\TopFrame{\gstack}$.
Finally, a configuration keeps track of a set of used provenances,
$\gused$ in order to ensure that fresh provenances can be assigned to
new allocations.

\subsection{The Specification Monad}
Memory models for compiler IRs are naturally nondeterministic
semantic objects: they must describe \emph{all} legal implementations
architectures may commit to, and allocations, in particular, are left
unconstrained, leading to nondeterminism.  We provide a specification
for each operation via a \textit{specification monad}:
\[\MemPropT{X} \defeq \MemState \to \SET{\erruboom{\MemState \times
      X}} \ \mathrm{where}\ \erruboom{A} \defeq \UB + \OOM + \FAIL +
  \ok{A}\]

\MemPropT{X} is stateful and nondeterministic, relating initial configurations
to a set of possible configurations that could result from executing a memory
operation. These sets are specified in a propositional way; we describe them
either via inference rules, or by composing them via the usual \tcinline{ret}
and \tcinline{bind} monadic operations.  Finally, the result type,
\erruboom{}, allows us to characterize the four possible acceptable behaviors of
an operation. We write $\isbeh{\var{c}~\varc{}}{\var{beh}}{}$ to state that
$\var{beh}$ is a valid behavior of a memory specification $\var{c}$ at initial
configuration $\varc{}$.

An operation may simply succeed, yielding \ok{\varc,\var{x}}, returning a new configuration \varc{} and
a resulting value $\var{x}$.  It may also raise one of three exceptional behaviors. The first is \emph{Undefined
  Behavior}, \UB{}, which arises from run-time situations that invalidate assumptions the compiler makes
to justify optimizations; semantically, these are modeled as computations that can
be refined into anything.
The second is an out of memory exception, \OOM{}.
This behavior captures all the situations in which the computation
may preemptively halt as a consequence of the representation of addresses; semantically it is modeled as a behavior that refines anything.
Lastly, operations may \FAIL, representing cases in our semantics that we intend to be
statically checked and ruled out. This exception also corresponds to language features that have not been implemented in our model. 
In the remainder of the paper, we therefore elide this case by providing
partial specifications instead. A key distinction between failure and \UB{} is that failure is not ``time-traveling'' in our semantics.

The monadic specification is particularly useful for maintaining very similar
structures between the memory model and the executable interpreter
(See Section~\ref{sec:memory-model-executable}), which simplifies
maintenance and the proof that the executable implementations of the memory model
respect the specifications.

\subsection{Low level memory operations}

\begin{figure}
    \begin{minipage}[l]{0.48\textwidth}
      \[
        \hspace{-0.5cm}
        \small{\begin{array}{{rcl}}
          \readbs{(\varptr : \Ptr)}{} & : & \MemPropT{\SByte} \\
                 \writebs{(\varptr : \Ptr)}{(\varb: \SByte)}{} & : & \MemPropT{\unitn} \\
          \pushfbs{} & : & \MemPropT{\unitn} \\
          \popfbs{} & : & \MemPropT{\unitn}
    \end{array}}
  \]
    \end{minipage}
    \begin{minipage}[r]{0.48\textwidth}
      \[
        \small{\begin{array}{{rcl}}
        \allocabs{(\varbs: \tlist{\SByte})} & : & \MemPropT{\Ptr} \\
      \mallocbs{(\varbs: \tlist{\SByte})} & : & \MemPropT{\Ptr} \\
      \freebs{(\varptr: \Ptr)}{} & : & \MemPropT{\unitn}
   \end{array}}
  \]
\end{minipage}
\vspace{-3ex}
  \caption{Memory model: low level operations}
  \label{fig:memory-interface}
\end{figure}

Signatures of the low level primitives interacting with the memory
model are described in Figure~\ref{fig:memory-interface}. The
operations are reads and writes of single bytes, allocations of blocks
(a contiguous sequence of bytes) on the stack and heap, operations for
freeing heap allocated blocks, pushing stack frames, and popping the
most recent stack frame in order to free stack allocated blocks.

\begin{figure}
\begin{mathpar}
  \inferrule{\noaccess{\varptr}{\gmem}}{\isbeh{\readbs{\varptr}{\varc}}{\UB{}}{\SByte}}\and
  \inferrule{\readbyte{\gmem}{\varptr}{\varb}}{\isbeh{\readbs{\varptr}{\varc}}{\ok{\varc,\varb}}{\SByte}}\and
  \inferrule{\noaccess{\varptr}{\gmem}}{\isbeh{\writebs{\varptr}{\varb}{\varc}}{\UB{}}{\unitn}}\and
  \inferrule
  {\preserveheap \quad \preservestack \quad \preserveused\\
    \access{\varptr}{\gmemo} \quad \readbyte{\gmemt}{\varptr}{\varb}\\
    \memokout{\gmemo}{\gmemt}{\singleton{\varptr}}
  }
  {\isbeh{\writebs{\varptr}{\varb}{\varc_1}}{\ok{\varc_2,\unitv}}{\unitn}}\\
\end{mathpar}
\vspace{-8ex}
\caption{Memory model: byte-level read and write operations}
\label{fig:prim-read-write}
\end{figure}

\begin{figure}
  \begin{mathpar}
  \inferrule{\varptr\not \in \gused}{\isbeh{\freshp{\varc}}{\ok{\updconfseed{\varc}{\{\varptr\}\cup \gused},\varptr}}{\Provenance}} \and
  \inferrule{\forall \vari,~0 \leq \vari < \var{n},\gmem[\vara+\vari] = \None}{\isbeh{\findfreebk{\var{n}}{\varc}}{\ok{\varc,\vara}}{\Addr}} \and
  \inferrule{
    \preserveheap \quad
    \preserveused \quad
    \gstackt = \TopFrame{\gstacko} \uplus \varptrs :: \TailFrame{\gstacko}\\
    \length{\varptrs} = \length{\varbs} \quad\forall \vari, 0\leq \vari <\length{\varptrs},~\readbyte{\gmemt}{\varptrs[\vari]}{\varbs[\vari]}\\
    \memokout{\gmemo}{\gmemt}{\varptrs}
   }
  {\isbeh{\allocabspost{bs}{ptrs}{m_1}}{\ok{m_2,\unitv}}{\unitn}}\\
  \inferrule{
    \preservestack \quad \preserveused
    \quad \gheapt = \updmap{\gheapo}{\varptrs[0].\vara}{\varptrs}
    \\
    \length{\varptrs} = \length{\varbs} \quad
    \forall 0 \leq\vari<\length{\varptrs},~\readbyte{\gmemt}{\varptrs[\vari]}{\varbs[\vari]}\\
    \memokout{\gmemo}{\gmemt}{\varptrs}
  }
  {\isbeh{\mallocbspost{\varbs}{\varptrs}{\varc_1}}{\ok{\varc_2,\unitv}}{\unitn}}\and
  \inferrule
  {
    \preservestack \quad \preserveused\quad
    \gheapo[\varptr.\vara] = \Some~\varptrs \quad
    \gheapt = \removemap{\gheapo}{\singleton{\varptr.\vara}}\\
    \forall \varptr'\in \varptrs,~(\access{\varptr'}{\gmemo} \land ~\gmemt[\varptr'.\vara] = None)\\
    \memokout{\gmemo}{\gmemt}{\varptrs}
  }
  {\isbeh{\freebs{\varptr}{\varc_1}}{\ok{\varc_2,\unitv}}{\unitn}}
\end{mathpar}
\vspace{-3ex}
\caption{Memory model: memory management primitives}
\label{fig:prim-mem-man}
\end{figure}

\begin{figure}
  \begin{mathpar}
    \inferrule{
      \preserveheap \quad
      \preserveused \quad
      \preservemem \quad
      \gstackt = \emptyset::\gstacko}{\isbeh{\pushfbs{\varc_1}}{\ok{\varc_2,\unitv}}{\unitn}}\and
    \inferrule{
      \preserveheap \quad
      \preserveused \quad
      \gstacko = \varptrs::\gstackt\\
     \forall \varptr\in \varptrs,~(\access{\varptr}{\gmemo} \land ~\gmemt[\varptr.\vara] = None)\quad
     \memokout{\gmemo}{\gmemt}{\varptrs}
   }{\isbeh{\popfbs{\varc_1}}{\ok{\varc_2,\unitv}}{\unitn}}
 \end{mathpar}
 \vspace{-3ex}
  \caption{Memory model: frame stack management}
  \label{fig:prim-stack-man}
\end{figure}

\paragraph{Out Of Memory behavior}
Perhaps surprisingly we \emph{always} allow operations to halt
preemptively with an out-of-memory behavior. One of our goals was for
the specifications to encompass a large number of possible
implementations for memory, and we've seen instances of memory models
that might ``run out of memory'' in counter-intuitive situations (for
instance a quasi-concrete memory model with a finite concrete memory
may run out of memory when a concrete block is allocated for a
pointer--integer cast), as such we've been very lenient with
allowing out-of-memory behaviors throughout or semantics. One could
tighten the specifications if desired, though when comparing the
infinite and finite memory models in
Section~\ref{subsec:relating-inf-fin-memory} our refinement relations
allow for out-of-memory anywhere in the finite memory model
anyway. Since out-of-memory is omnipresent, we work under the
convention that all specifications $c:\MemPropT{X}$ may run out of
memory, i.e., $\isbeh{c~m}{\OOM{}}{X}$ is satisfied for any initial
memory $m$. This makes \OOM{} a kind of refinement dual to \UB{}. While \UB{} can
always be refined into any computation, any computation may be refined
by \OOM{}, as we purposefully do not want to reason about programs that
run out of memory.

\paragraph{Reading and writing bytes (Figure~\ref{fig:prim-read-write})}
The operation $\readbs{\varptr}{\varc}$ specifies the possible behaviors when dereferencing a pointer
in memory.
Dereferencing boils down to looking up the memory, with the additional
provenance check introduced in Section~\ref{sec:configurations}.
If the lookup fails, or is illegal, \UB{} may be raised.
Writing a byte $\varb$ to memory $\gmemo$ at a pointer
\varptr{} may trigger \UB{} in similar cases to reading. A successful
write must furthermore specify the resulting configuration $\varc_2$---the statement
is made slightly more complex to account for provenance, but hides no surprise. Only the memory
component of the configuration is modified. A pointer $\varptr$ accessible in
$\gmemo$ will be remapped to the byte $\varb$ in $\gmemt$. Finally, all addresses distinct from
$\varptr.\vara$ are unchanged in memory.

\paragraph{Memory management (Figure~\ref{fig:prim-mem-man})}

Specifying allocation is a more involved and less atomic task, so we
leverage the specification monad to describe it in a more programmatic
style.  Utility specifications help generate a fresh provenance \varpr
($\mathtt{fresh}$), and retrieve an available contiguous sequence of
$\var{n}$ addresses in memory
($\mathtt{find\_bk}~\var{n}$). \footnote{In line with LLVM, allocating
  no bytes (i.e. an empty list) may return \textit{any} address. Its
  fresh provenance, which no data is equipped with, will, however,
  ensure we cannot do anything with this address without triggering
  \UB.} Note that $\mathtt{find\_bk}$ uses $a+i$ to compute contiguous
addresses, should this overflow in the finite model the allowed
behavior will only be $\OOM$.

Stack and heap allocations of a list of bytes are specified very similarly,
retrieving a fresh provenance, an available range of addresses, constraining the
resulting memory, and finally returning the first allocated address:

\begin{minipage}[l]{0.48\textwidth}
  \hspace{-0.5cm}
  \small{$\begin{array}{ll}
            \multicolumn{2}{l}{\allocabs{(\varbs:\tlist{\SByte})}   : \MemPropT{\Ptr} := } \\
            \enskip & \varpr \leftarrow \freshp{} \mbox{;;} \\
                  & \vara \leftarrow \findfreebk{(\length{\varbs})}{} \mbox{;;} \\
                  & \allocabspost{\varbs}{[(\vara,\varpr),\dots,(\vara+\length{\varbs}-1,\varpr)]}{} \mbox{;;} \\
                  & \tcinline{ret}\ (\vara,\varpr)
          \end{array}$}
      \end{minipage}
      \begin{minipage}[r]{0.48\textwidth}
        \small{
          $\begin{array}{ll}
             \multicolumn{2}{l}{\mallocbs{(\varbs:\tlist{\SByte})}   : \MemPropT{\Ptr} := } \\
             \enskip & \varpr \leftarrow \freshp{} \mbox{;;} \\
                   & \var{a} \leftarrow \findfreebk{(\length{\varbs})}{} \mbox{;;} \\
                   & \mallocbspost{\varbs}{[(\vara,\varpr),\dots,(\vara+\length{\varbs}-1,\varpr)]}{} \mbox{;;} \\
                     & \tcinline{ret}\ (\vara,\varpr)
           \end{array}$}
       \end{minipage}

\noindent These specifications only differ in how the configurations are constrained as depicted on
Figure~\ref{fig:prim-mem-man}.
For stack allocation, $\allocabspost{\varbs}{\varptrs}{}$
ensures that (1) the new pointers are added to
the current stack frame, (2) the addresses are written with the corresponding
bytes, all sharing the provenance \varpr, and (3) nothing else in memory is altered.
For heap allocation, $\mallocbspost{\varbs}{\varptrs}{}$ enforces (2) and (3)
similarly, but, instead of manipulating the stack, it ensures that
the returned address is a root in the heap associated with
the set of newly allocated pointers.

Given a pointer \varptr, $\freebs{\varptr}{\varc_1}$ ensures that
$\varptr.\vara$ is a valid root in $\gmemo$, associated with a set of pointers
$\varptrs$ accessible in $\gmemo$.  Under this assumption, it simply reclaims
the pointers and severs $\varptr.\vara$ from the heap. Though not shown in the
Figure, UB occurs in the complementary cases---when the pointer is not a root in
the heap or if the block was not actually allocated in memory.

\paragraph{Stack management (Fig~\ref{fig:prim-stack-man})}
Pushing a new frame, $\pushfbs{}$, is trivial, simply adding the
emptyset on top of the stack. The specification of $\popfbs{}$ is
very close to $\mathtt{free}$: we ensure all pointers in the top
frame are accessible in the original memory, reclaim them, and pop the
stack.

\subsection{Relating the Infinite and Finite Memory Models}
\label{subsec:relating-inf-fin-memory}

We now consider two instances of the memory model, based on two
representations of addresses: an \emph{infinite} memory model with
unbounded integers, and a \emph{finite} one with 64-bit integers. We
will distinguish between these models with \emph{inf} and \emph{fin}
superscripts for the infinite and finite memory models respectively.

Our overarching goal is to be able to reason about programs in the
infinite memory model, performing program transformations under the
infinite semantics. We will then convert these infinite memory
programs to finite memory programs that will eventually be compiled
to native assembly code for a concrete architecture.  In order to
ensure that this process is sound, we must relate the behavior of
memory operations under the infinite model with the behavior of the
same operations under the finite model.

The rough idea is to consider the execution of programs under the
infinite model when their allocations happen to fit within the finite
memory model's range of memory addresses---executions that don't fit will
exhibit \OOM behavior. We then ensure that operations on these finite
memory slices agree between the infinite and finite memory
models. Essentially, the finite memory model should be able to simulate
the infinite memory model, as long as all the addresses stay in
bounds!

\subsubsection{Relating Configurations}

\newcommand{\varcinf}{\INF{\varc}}
\newcommand{\varcfin}{\FIN{\varc}}

\begin{figure}
\[
  \begin{array}{ccll}
    \varcinf{} \refineIF \varcfin{} & := & \varcinf{} = \lift{\varcfin{}} & \mbox{Infinite to finite refinement} \\
\\
    \lift{\varcfin{}} & := &
                          \left\{\begin{array}{lcl}
                                   \fmem & = & \{ z \mapsto (\lift{b}, p) \ | \ \gmemg{\varcfin{}}[z] = (b, p) \} \\
                                   \fheap & = & \{ z \mapsto \mathtt{map}\; \lift{-}\;\overline{\var{blk}} \ | \ \gheapg{\varcfin{}}[z] = \overline{\var{blk}} \} \\
                                   \fstack & = & \mathtt{map}\; (\lambda \overline{\var{blk}}
                                                 \cdot \mathtt{map}\;\lift{-} \overline{\var{blk}})\; \gstackg{\varcfin{}}  \\
                                   \fused & = & \gusedg{\varcfin{}} \\
                                 \end{array}\right. & \mbox{Configuration Lifting}\\
  \end{array}
\]
\[
  \begin{array}{ccll}
    \lift{\varptr}      & := & (\IntToZ{\gadd},\gpr) \\
    \lift{b}         & := & \multicolumn{2}{l}{\liftsbyte{(b)} \quad
                            \mbox{(\textit{lifting symbolic bytes, language specific})}}
  \end{array}
\]
\vspace{-3ex}
\caption{Infinite-to-finite refinement defined by lifting of finite
  memory configurations into infinite memory configuration. Here, \liftsbyte{} is
  a language-specific lifting of finite $\SByte^{\var{fin}}$ into
  $\SByte^{\var{inf}}$. We omit implicit $\IntToZ{}$ casts when the integer is
  known to be within finite bounds.}
\label{fig:inf-to-fin-refinement}
\end{figure}

We start by relating configurations at both levels.
As described in Figure~\ref{fig:inf-to-fin-refinement},
a finite $\varcfin{} : \MemStateFin{}$ is a refinement of an infinite
$\varcinf{}$, written $\varcinf{} \refineIF \varcfin{}$, when $\varcinf{}$
coincides with the lifting $\lift{\varcfin{}}$.
Lifting a \MemStateFin{} is fairly straightforward, both domains of
configurations having similar concrete representations. The lifting therefore
simply maps over the structure the trivial injection of finite addresses into
$\mathbb{Z}$, as well as the lifting of symbolic bytes.\footnote{We delay this
  description to Section~\ref{subsubsec:sbytes}.}

\subsubsection{Relating Operations}

\newcommand{\binf}{\ensuremath{b^{\var{inf}}}}
\newcommand{\bfin}{\ensuremath{b^{\var{fin}}}}

We can now capture the intuitive expected behavior of the memory operations: if the memory configuration
can fit in the finite representation, then the same behavior can be observed.
In practice, we prove a refinement lemma for each low-level operation. For
instance, in the case of reads:

\begin{lemma}[$\mathtt{read\_byte\_spec}$ refinement]\  \\
  If \ \ 
  \begin{minipage}[c]{2.2in}
    \noindent
    \begin{itemize}[leftmargin=*]
    \item  $\varcinf{} \refineIF \varcfin{}$, and
    \item  $\ptrinf{} = \lift{\ptrfin{}}$, and
    \item  $\isbeh{\readbs{\varcinf{}}{\ptrinf}}{\ok{\varcinf,\binf}}{}$
    \end{itemize}
  \end{minipage}
  then  $\exists\; \bfin{}$ such that \ 
  \begin{minipage}[c]{2.5in}
    \noindent
  \begin{itemize}[leftmargin=*]
  \item  $\binf = \lift{\bfin}$, and
  \item  $\isbeh{\readbs{\varcfin{}}{\ptrfin}}{\ok{\varcfin,\bfin}}{}$.
  \end{itemize}
  \end{minipage}
\end{lemma}

The lemmas for the other operations are similar in shape, so we omit them for conciseness.

%% file: integrating-into-llvm.tex
The infinite and finite models above provide general semantics that
are parameterized by addresses and symbolic bytes, as
these parameters could vary depending on the programming language. To
ensure that it is sufficiently expressive for use in practice, we have
instantiated the framework in the context of an existing formal
semantics, \vir~\cite{zakowski2021}.  \vir aims to be a specification
for a large, practically applicable subset of (sequential) LLVM IR. It
supports a rich, C-like structured memory model, including
integer--pointer casts; it also includes \cinline{undef} and
\cinline{poison} values that interact non-trivially with the memory
model specification because they affect the notion of UB. The prior
\vir memory models were based on the CompCert  and quasi-concrete
semantics, and so they suffered from the deficiencies with respect to
optimization correctness mentioned in
Section~\ref{sec:remembering-memory-models}. The new memory model
rectifies those problems and is also more faithful to LLVM IR's intended
semantics with respect to \cinline{undef}.  This Section describes the
\vir instantiation of the framework and, along the way, addresses
some challenges of formalizing LLVM IR semantics. As in the general
framework, this instantiation yields both an \textit{infinite memory}
and a \textit{finite memory} version of the \vir semantics.  Most of
the development is parametric with respect to that choice, but we
differentiate them as \INF{\vir} and \FIN{\vir} where necessary.

\subsection{Layered interpreters}

\begin{wrapfigure}{r}{3in}
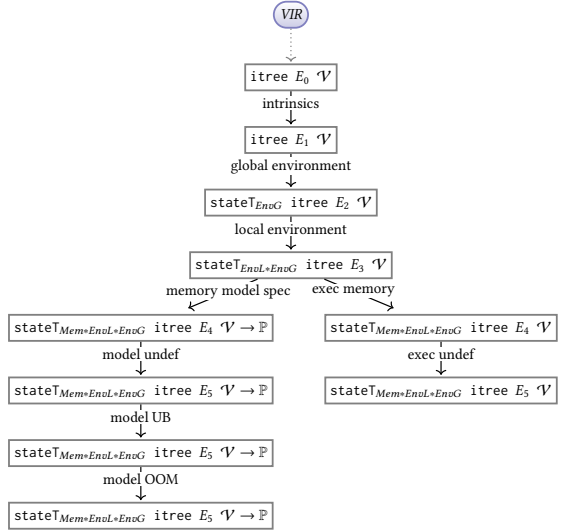


    \usetikzlibrary {arrows.meta,graphs,shapes.misc,quotes}
    \resizebox{2.9in}{!}{\tikz [black, text=black, thick,
    every new ->/.style = {shorten >=1pt},
    language/.style = {
      rounded rectangle, minimum size=6mm, very thick, draw=blue!50!black!50, top color=white,
      bottom color=blue!50!black!20, font=\itshape, text height=1.5ex,text depth=.25ex},
    interpreter/.style = {
      rectangle,  minimum size=6mm, very thick, draw=black!50, font=\ttfamily, text height=1.5ex, text depth=.25ex}  ]
    \graph [
    edge quotes={fill=white,inner sep=1pt, pos=0.35},
    grow down sep=8mm, branch left=70mm, simple] {
      VIR[language] ->[black!50, dotted]
      L1/{\itreef{E_0}~\dvalue} [interpreter] ->["intrinsics"]
      L2/{\itreef{E_1}~\dvalue} [interpreter] ->["global environment"]
      L3/{\stateT{\envG}{\itreef{E_2}}~\dvalue} [interpreter]  ->["local environment"]
      L4/{\stateT{\envL * \envG}{\itreef{E_3}}~\dvalue} [interpreter] ->[shorten >=10pt]
      {[nodes={xshift=35mm}]
        L5EXEC/{$\stateT{\mem * \envL * \envG}{\itreef{E_4}}~\dvalue$} [interpreter, > "exec memory"] ->["exec undef"]
        L6EXEC/{$\stateT{\mem * \envL * \envG}{\itreef{E_5}}~\dvalue$} [interpreter],
        L5PROP/{$\stateT{\mem * \envL * \envG}{\itreef{E_4}}~\dvalue \rightarrow \PROP$} [interpreter, > "memory model spec"] ->["model undef"]
        L6PROP/{$\stateT{\mem * \envL * \envG}{\itreef{E_5}}~\dvalue \rightarrow \PROP$} [interpreter] ->["model UB"]
        L7PROP/{$\stateT{\mem * \envL * \envG}{\itreef{E_5}}~\dvalue \rightarrow \PROP$} [interpreter] ->["model OOM"]
        L8PROP/{$\stateT{\mem * \envL * \envG}{\itreef{E_5}}~\dvalue \rightarrow \PROP$} [interpreter]
      };
    };
  }
    \caption{Levels of interpretation. Each box shows the type of the semantic definitions at that layer.  Arrows are labeled with the events they define.}
    \label{fig:interp}
  \end{wrapfigure}

\vir is structured as a series of \textit{layered interpreters}, as shown in
Figure~\ref{fig:interp}, each of which specifies some aspects of the LLVM IR
semantics. These interpreters are built on top of \textit{interaction
  trees}~\cite{itrees,yzz22}, which are a Coq datatype of potentially infinite trees
(used to model diverging programs) whose nodes are \textit{uninterpreted}
events, indicated by $E_0$--$E_5$ in the Figure. Each layer \textit{handles}
some subset of the events, defining their semantics, and leaving the rest for
later layers to handle. 

For the purposes of this paper, the most important parts of the interpretation
stack are the handlers for \textit{memory events}, \textit{nondeterminism},
\textit{undefined behavior}, and \textit{out-of-memory} exceptions. The memory events,
\memE, correspond to LLVM IR operations that interact with the memory model and
each of those events is parameterized by appropriate input values and a return
type \dvalue or \uvalue{} (indicated as a superscript), as shown below---these types, describing \textit{dynamic values}, are explained in the next Subsection.  The
other kinds of events are similarly annotated.\footnote{Note that the
  superscript $\emptyset$ for \ub and \oom means that these events do not return
  any value; they simply terminate the computation. The \picku{uv} and \picknp{uv} events
  technically return \textit{constrained} types that guarantee uniqueness or a non-\dvpoison value, respectively, but we elide those details here.
}

\[\small
  \renewcommand{\arraystretch}{0.7}
  \begin{array}{rclrcl}
    \multicolumn{3}{l}{\mbox{Memory model interaction events}} & \multicolumn{3}{l}{\mbox{Undefined Behavior event}} \\
    \memE & \defeq & \mpush\vor\mpop\vor\load{\tau}{a}\vor\store{a}{uv}  & \ubE & \defeq & \ub\\
          & \vor   & \alloca{\tau} \\ \\
    \multicolumn{3}{l}{\mbox{Nondeterminism events}} & \multicolumn{3}{l}{\mbox{OOM event}} \\
    \pickE & \defeq & \pick{uv} \vor \picku{uv} \vor \picknp{uv} & \oomE & \defeq & \oom \\
  \end{array}
\]

With respect to the memory model, the main work of a \vir semantics is to
implement a handler for the \memE{} events in terms of the primitive operations
described in Section~\ref{sec:memory-model-details}.  That handler is plugged
into the stack of Figure~\ref{fig:interp} to interpret $\memE \in E_3$ into
stateful operations that manipulate the memory.  As shown in the left-hand-path
of the interpretation stack, the \textit{specification} of the memory model is
defined propositionally in Coq to account for the nondeterminism of possible
implementations, including nondeterminism introduced by the memory model. The right-hand
path implements the \textit{executable} version of the semantics.
On the specification side, the handlers for \ubE{}, \pickE{} and \oomE{} events interact
with the notion of \textit{refinement} for the behaviors of \vir{} programs (see
Section~\ref{sec:vir-refinement}).  There are no equivalent handlers for \ubE{}
and \oomE{} on the executable side, because the OCaml framework for executing
\vir \itreesn simply raises an exception when those events are encountered. 

If $p$ is a \vir program, we write, e.g., $\sem{p}_3$ to mean interpretation
down to the $3^{rd}$ layer (i.e., just before the split). We write
$\sem{p}_{\vir}$ for the ``fully interpreted'' semantics of $p$, i.e., at the
bottom of the left branch, and we write $\tm{interpret}\; p$ for the executable
version, at the bottom right branch.  That the executable semantics is a valid
implementation of the specification is one of the main theorems about the \vir
development (see Theorem~\ref{thm:interpreter-is-sound}).

\subsection{\vir{} Values}
\label{subsec:vir-values}

\subsubsection{Dynamic values and \intptr}

The semantics of \vir relies upon the domain \dvalue of \textit{dynamic values} that
the language can manipulate. The core of these dynamic values are the
\emph{defined values}.
{\small
$$dv\in\dvalue ::=~ \dvnone\sep\dvaddr{a}\sep\dvarray{\tlist{\dvalue}}\sep\dvint{i}{1}\sep\dvint{i}{8}\sep\dvint{i}{16}\sep\dvint{i}{32}\sep\dvint{i}{64}\sep\dvinfi{i}\sep \dvpoison_{\tau}$$
}%
The void value, \dvnone, is a placeholder for operations with no meaningful
return values.  Memory addresses (\dvaddr{a}), of type \Addr{}, are implemented
either as positive integers, $\INF{\Addr{}} = \mathbb{Z}$, or 64-bit values $\FIN{Addr{}} = \tm{i64}$, depending on whether we are
instantiating the infinite or finite memory model.
\vir supports all of \ir's structured values, including records, but here we present only arrays, noted as \dvarray{$dv_1,\ldots,dv_n$}.

\vir supports \tm{i1}, \tm{i8}, \tm{i16}, \tm{i32} and \tm{i64}-bit integers\footnote{We
  use CompCert's finite integers in our development. \vir also supports floating-point values (omitted here for simplicity).} ranged over by
\dvint{i}{1}, \dvint{i}{8}, etc., but it also includes integers of type \intptr{}, which are in bijection with memory addresses, \footnote{The \intptr{} type is 
  missing from the LLVM IR, but there is precedent for it in C, via
   \cinline{intptr_t}, \cinline{uintptr_t} and \cinline{size_t}
  \cite{c99standard}} and ranged over by \dvinfi{i} .  \intptr{} has the same
cardinality as the \Addr{} type, i.e., $\INF{\intptr} \defeq \mathbb{Z}$ and
$\FIN{\intptr} \defeq \var{i64}$.
The \intptr{} type acts mostly like the other
integer types, supporting all of the same instructions, which allows
for programs to perform arbitrary arithmetic on physical addresses
\emph{without} forcing a cast to a type of a fixed finite size. Because \intptr{} has the same cardinality as \Addr{},
pointer--integer casts can effectively be pure no-ops, allowing
these casts to be removed to make way for other optimizations, and
ensuring a round-trip property where a pointer cast to an
integer and back yields the same pointer\footnote{Note that
  there are concerns about provenance for such casts that we do not
  tackle in this paper~\cite{VIP}. Our implementation does not track provenance
  through integers, so our integer--pointer casts will always yield
  a pointer with a wildcard provenance, which is safe, but can limit
  optimizations in rare cases (see the discussion in
  \cite{memarian2019exploring}). This provenance tracking is largely
  orthogonal to the design of our memory model, however, our memory
  model can be integrated with a language regardless of whether it
  tracks provenance through integer computations.}. In order to preserve this round-trip property for casts,
arithmetic on \intptr{} values in \FIN{\vir} may provoke \OOM, as discussed in Section~\ref{sec:arithmetic}.

\vir also includes \emph{poisoned values} (\dvpoison)
representing a \emph{deferred} undefined behavior~\cite{LKSH+17}. Deferred UB is
instrumental for aggressive optimizations, but a semantic subtlety.  The
\dvpoison value is a tainting mark: it propagates to all values that depend on
it, so equivalences such as
$\dvpoison + \dvpoison \equiv 2 * \dvpoison \equiv \dvpoison$ hold true.

\subsubsection{Undef Values and Symbolic Bytes}
\label{sec:handling-undef}
\label{subsubsec:sbytes}

LLVM represents uninitialized memory through \textit{undefined values}, which
represent the set of possible values of a given type, and operations on them
manipulate a those sets.  Reasoning about undefined values is subtle; each time
an \undefv{} value is used within an LLVM program, it may take on a different
concrete value.  For instance,
$\uvundef{\isf} + \uvundef{\isf} \equiv \uvundef{\isf}$, whereas
$\uvundef{\isf} + \uvundef{\isf} \not \equiv 2 * \uvundef{\isf}$,
because the value on the right hand side of the equation cannot be an odd
number.

In \vir, ``\textit{uvalues}'' or \emph{under-defined values}, written $\textit{uv} : \uvalue$,
model these sets. They are given by:

{\small
  \[
  \begin{array}{l}    
    \textit{uv}\in\uvalue ::=~ \dtou{\dvalue}\sep [\tlist \uvalue]\sep\uvundef{\tau}\sep\tm{op}~ \uvalue~\uvalue\sep [\tlist \sbyteT]_{\tau}
    \\
    \textit{sb}\in\sbyteT ::=~ \uvsbyte{\tau}{uv}{i}{sid}
  \end{array}
  \]
}%

Under-defined values include defined (i.e. concrete) values---we write
$\uparrow$ for the corresponding injection---as well as \textit{arrays} of
uvalues.  They also include \uvundef{\tau}, which stands for the \textit{set} of
all possible concrete values of the type $\tau$ (we omit $\tau$ when it is
unimportant). \uvalue{} also includes ``delayed'' operations, where \tm{op}
ranges over \vir{}'s arithmetic, bit-logic, and other computation primitives.
Such a uvalue lifts the set semantics of \uvundef{} to nondeterministic
computations, as we explain below.  The uvalues also contain (type-annotated)
concatenations of \textit{symbolic bytes}~\cite{besson2019compcerts},
$\textit{sb} : \sbyteT{}$, explained next.

\paragraph{Serializing under-defined values}
Values in \vir is stored to memory as list of bytes.  An undefined value is
serialized as symbolic bytes, given by the type \sbyteT{}.  A
\uvsbyte{\tau}{uv}{i}{sid} represents the $i^{th}$ byte of the value $uv$ with a
store-ID $sid$, whereas $[\tlist \sbyteT{}]_{\tau}$ concatenates a series of symbolic
bytes into a under-defined value of type $\tau$.

To enable optimizations like store-forwarding, the semantics must precisely
preserve nondeterminism when serializing and deserializing \uvaluesn to and from
bytes. For example, the event handler for \tm{Store} serializes a
$uv \in \uvalue$ into an array of symbolic bytes matching the size of the type
(written $|\tau|$), with each byte containing the appropriate index into the
\uvaluen. This serialization operation is defined as:
{\small
\[
 \begin{array}{rcl}
   \tm{serialize}(uv,\tau) & = & sid \leftarrow \tm{fresh\_sid} ;; \\
                           &   & \kw{ret}\ [\uvsbyte{\tau}{uv}{0}{sid},\ldots,\uvsbyte{\tau}{uv}{|\tau|}{sid}]_{\tau}
\end{array}
\]  
}%
Each of these symbolic bytes contains that same ``store id'' ($sid$), which is
uniquely generated for every \tm{Store} event, to preserve ``entangled''
\undefv{} values within the semantics.  This $sid$ is assigned to the serialized
bytes and it is used to prevent the introduction of too much nondeterminism when
reading bytes written by multiple stores.  For example, the program in
Figure~\ref{fig:entangled-bytes} illustrates the scenario in which the first two
bytes and final two bytes of \cinline{z} are entangled together, so there are
only two possible values for each of these two byte chunks.  Note that, if two
symbolic bytes have the same $sid$, they must have come from the same store, and
thus agree on their underlying $uv$ too.

\begin{figure}
  \begin{noerr}
\begin{minted}[]{llvm}
%x = i16 select i1 undef, i16 0x1234, i16 0x0000
%y = i16 select i1 undef, i16 0x5678, i16 0x0000
store i16 %x, ptr %ptr
store i16 %y, ptr (inttoptr (add iptr (ptrtoint %ptr) 2)
%z = load i32, ptr %ptr
\end{minted}
\end{noerr}
\vspace{-3ex}
  \caption{Entangled bytes. Here the \tm{select} instruction chooses between two values nondeterministically due to \tm{undef}, but the \tm{load} should read (assuming big-endianess for simplicity) one of
\cinline{0x00000000}, \cinline{0x00005678}, \cinline{0x12340000}, or
\cinline{0x12345678}, but never something like \cinline{0x00345600}.}
\label{fig:entangled-bytes}
\vspace{-3ex}
\end{figure}

Conversely, the handler for the \tm{Load} instruction \textit{deserializes} the
symbolic bytes:
{\small
\[
  \begin{array}{l}
    \tm{deserialize}([sb_1,\ldots,sb_n]_{\tau}, \tau') \ =  \\
    \kw{if}\ \tau = \tau' \AND [sb_1,\ldots,sb_n]_{\tau} = [\uvsbyte{\tau}{uv}{0}{sid},\ldots,\uvsbyte{\tau}{uv}{|\tau|}{sid}]_{\tau}\ \kw{then} \ \ uv \ \ \kw{else} \ \ [sb_1,\ldots,sb_n]_{\tau'}
  \end{array}
\]  
}%
In the simple case where the value is loaded with the same type, the
deserialization for \tm{Load} can simply extract the original $uv$\footnote{There is a corner case for types where $|\tau| = 0$ (such as an array of length zero); in that case the correct $uv$ is uniquely determined by $\tau$.} (a property
which makes store forwarding optimizations easy to justify). In more complex
cases, where bytes are read at a different type from what they are stored at
(possibly reading a portion of the bytes from
different \uvaluesn) the resulting \uvaluen is left as the concatenation of
symbolic bytes but with the updated type---their concrete bit patterns will be resolved via
``concretization,'' as explained next.

\subsection{Concretization: Refinement and Evaluation of LLVM Values}
\label{sec:concretization}

As we saw above, an under-defined value $uv$ denotes a \textit{set} of
``concrete'' dynamic values---that is, it is a \textit{specification} of the
set of allowed bit patterns a compliant implementation can use to refine $uv$.
We write $\concretize{-} : \uvalue \to \SET{\erruboom{\dvalue}}$ to denote the
(monadic) function that computes a set of concretizations of $uv$.  We write
$\uvrefine{dv}{uv}$, defined as $\concretize{uv} \ni \ok{dv}$ to indicate that
(concrete) dynamic value $dv$ is a legal \textit{refinement} of $uv$.  The
concretization function is implemented in a similar fashion to the prior \vir
semantics \cite{zakowski2021}. It essentially implements an interpreter for
all of the computational instructions ``lifting'' them to work on sets of
values. One important base case is that \tm{\uvundef{\tau}} concretizes to the
set of all legal values of type $\tau$, that is:
$\concretize{\uvundef{\tau}} = \{dv | 
dv : \tau, dv \not = \dvpoison
\}$. For example, we have \uvrefine{\tm{2}}{\tm{mul\
    i64\ 2,\ i64\ 1}} and also \uvrefine{\tm{2}}{\tm{mul\ i64\ 1,\
    \uvundef{\tm{i64}}}} but \ensuremath{\tm{2} \not\in \concretize{\tm{mul\
      i64\ 3,\ \uvundef{\tm{i64}}}}}.  As you can see, due to the presence of
arithmetic (and other non-trivial) LLVM IR operations, and the fact that
under-defined values \textit{include} ordinary values as a special case, the
refinement relationship acts as an \textit{evaluation} relation---in the case
that $uv$ has no occurrences of \tm{undef} (i.e., it is defined), then
\uvrefine{dv}{uv} simply means that $uv$ evaluates to $dv$ according to the
ordinary rules of LLVM IR computations, as in the first example
above.

Note that concretization can fail with \UB (in case of, for example, division
by \tm{0}) or with \OOM (when working with \tm{iptr} values, as described below).

\subsubsection{Concretizing symbolic bytes}

New to this work is the treatment of symbolic bytes. Recall that symbolic
bytes represent byte-sized fragments of (possibly) under-defined values and
that a \tm{Load} event might read a sequence of such bytes that were written
by  (several) different \tm{Store}s.
Series of symbolic bytes are concretized as shown below:
\[
  \begin{array}{l}
    \concretize{[\uvsbyte{\tau_1}{uv_1}{0}{sid_1},\ldots,\uvsbyte{\tau_n}{uv_n}{n}{sid_n}]_{\tau}} = \\
    \quad dv_0 \leftarrow \concretize{uv_0}\; \mbox{;;}\; \ldots \; \mbox{;;}  \; dv_n \leftarrow \concretize{uv_n} \; \mbox{;;} \\
    \quad \{ \tm{ok}(\bitcast{\tau}{dv_1[1]dv_2[2]\ldots{}dv_n[n]}) \ | \ \forall j k,\; sid_j = sid_k \IMPLIES dv_j = dv_k \}
  \end{array}
\]
This works by recursively concretizing the $uv_i$'s, each of which yields a
set of concrete dynamic values \concretize{uv_i} (or an error, in which case
the whole concretization is an error).  Then, from each
$dv_i \in \concretize{uv_i}$ we can extract the (concrete) $i^{th}$ byte,
written as $dv_i[i]$.  The resulting set of concrete values is then obtained
by concatenating the individual bytes combinatorially; however, if two
symbolic bytes share a $sid$ (and hence come from the \textit{same}
\tm{store}) they are ``entangled'' and must be concretized in the same way.

The resulting sequence of bytes is converted to a dynamic value of type $\tau$
by the \bitcast{\tau}{-} operation. It might need to truncate or pad the
sequence, depending on the size of values of type $\tau$ and the number, $n$,
of available bytes.  This \tm{bitcast} operation is, ultimately, what allows
the conversion between values of distinct types.  For instance, it could
convert an array value of type $[8 \times \tm{i8}]$ into a value of type
\tm{i64}, even though such values have different representations in the
semantics. Altogether, this treatment of symbolic bytes properly ensures
``entanglement'' of values as illustrated in Figure~\ref{fig:entangled-bytes}.

\subsubsection{Concretizing \INF{\intptr{}} values} For the infinite memory
model, in which \intptr{} is taken to be (unbounded) integers, concretization
works in essentially the same fashion as the other integer types. This is the
easy case, because in \INF{\vir} \intptr{} values are just $\mathbb{Z}$
values, so no overflow or underflow can occur, and all of the operations work
straightforwardly as expected.

\subsubsection{Concretizing \FIN{\intptr{}} values}
\label{sec:arithmetic}
The finite memory model defines \intptr{} to be unsigned 64-bit integers.
However, unlike for ordinary \tm{i64} arithmetic operations, in which LLVM IR's
\cinline{nuw} and \cinline{nsw} (``No un/signed wrap'') flags cause
overlow/underflow to be treated as introducing undefined behavior, for
\intptr{}, such errors instead introduce \OOM.  This difference from
``ordinary'' integers is crucial to maintaining the connection between the
infinite and finite semantics.  To see why, consider the following program
(written using C-like notation that is easy to express as \vir code):

\begin{minted}[]{c}
iptr i = 1;
while (0 < i) { ++i; printf("%zd\n", i); }
do_evil();
\end{minted}

In the infinite language the \intptr{} addition can never overflow, so this
program will count up indefinitely, never calling the \cinline{do_evil}
function. If we na\"{i}vely ``convert'' this program to a finite program by
simply using 64-bit arithmetic, which can wrap, the value of \cinline{i} will
overflow to the value \cinline{0}, terminating the loop and thus calling
\cinline{do_evil}, which is not an allowed refinement.  From this example we can
see that it's clearly \emph{not} safe to allow \FIN{\intptr} values to wrap in
general, as that can change the meaning of the program. LLVM's \cinline{nuw}
flag also does the wrong thing---it introduces undefined behavior, so translating
the infinite program to a finite program in this way would cause the target
program to have UB while the source program does not!

Ultimately, the only reasonable solution is to add bounds checks to
integer operations on \intptr values and to halt the program with \OOM
when the checks fail (intuitively, such an arithmetic operation has
run out of bits in which to store the result). The \FIN{\vir}
semantics incorporates these bounds checks directly into the
specification of arithmetic on \intptr values, as part of
concretization, but these bounds checks could be added explicitly on
top of regular \tm{i64} values if desired.

  \subsection{Behavioral Refinement within $\vir^{X}$}
\label{sec:vir-refinement}

If we fix our attention on just one of \INF{\vir} or \FIN{\vir}---call it
$\vir^{X}$---and consider the interpretation stacks as shown in
Figure~\ref{fig:interp}, there are several notions of \textit{behavioral
  refinement} that are useful for reasoning about the semantics.  First, there
is refinement at each successive layer of interpretation---that is, we can think
of interpretation down to each layer as defining a program semantics with its
own notion of refinement. Following~\cite{zakowski2021}, a key result about of
the \vir development shows that refinement at one layer \textit{implies}
refinement at the next layer, which allows reasoning at one stage of the
interpretation stack to be used to prove results about the ``full'' semantic
interpretation. 

Up until the interpretation of memory events, refinement is built on
stateful variants of the $\eutttext_R$ bisimulation relation as
defined previously~\cite{zakowski2021}.  For instance, after
interpreting the local and global environments at layer 2, we would
have the following top-level refinement relation between the behaviors
of programs under a given environment represented by the itrees $P$
and $Q$ of type $\itree{E_2}{(\envL \times (\envG \times \dvalue))}$:
\[
  \langrefine{2}{Q}{P} \quad := \quad \eutttext_{\approx_{env_2}}(P,\ Q)
\]
Typically, we instantiate the refinement relation by using it on the
interpretations of program \textit{syntax}, i.e., by taking
$P = \sem{p}_2\;g\;l$ and $Q = \sem{q}_2\;g\;l$, where $g$ and $l$ are
global and local environments. The relation $\approx_{env_2}$ acts as
a postcondition on the results computed by the $P$ and $Q$; in this
case, it states that the returned values are equivalent, ignoring the global and local environments.

\newcommand{\hasub}[1]{\ensuremath{\tm{hasUB}(#1)}}
\newcommand{\oeutttext}{\tm{eutt\_oom}}

Once memory events are interpreted the semantics is nondeterministic,
as the handler for \memE{} events (which defines the meanings of
\tm{MPush}, \tm{MPop}, \tm{Load}, \tm{Store}, and \tm{Alloca}) is
implemented using the nondeterministic primitives of the general
memory model framework operations from
Figure~\ref{fig:memory-interface}, along with the \tm{serialize} and
\tm{deserialize} mechanisms described above. The interpretation of
\pickE{} events also introduces further nondeterminism due to the
treatment of \tm{undef} values. The refinement relation after
interpreting memory and pick events is given by a set inclusion
relation between sets of itrees $P$ and $Q$ of type $\SET{\itree{E_4}{\mem \times (\envL \times (\envG \times \dvalue))}}$:
\[
  \langrefine{4}{Q}{P} \quad := \quad \forall t' \in Q, \exists t \in P, \eutttext_{\approx_{env_4}}(t, t')
\]
The sets of itrees are generally taken to be those given by the
interpretation of program syntax using the propositional semantics, so
$P = \sem{p}_4\;g\;l\;sid\;m$ and $Q = \sem{q}_4\;g\;l\;sid\;m$, where
$g$ and $l$ are the initial global and local environments as before,
$sid$ is the initial high watermark for store ids, and $m$ is the
initial state of the memory.

Finally, we would like to take UB and OOM into account.
The semantics of \ub{} provides ``time traveling'' undefined behavior
semantics~\cite{timetravelub}. Intuitively, any program, here represented as an
interaction tree, that may reach a \ub{} event is considered to be ill-defined.
We write this predicate as \hasub{t}, and, in that case, \textit{any} other
behavior is allowed in its set of refinements.  Dually, the treatment of \oomE{}
events says that an \textit{out-of-memory} event refines \textit{any} behavior
(but not in a ``time-traveling'' fashion---the programs must agree up until the
\oom{} occurs).  That notion is defined via a modified version of the ordinary
$\eutttext{}_R$ relation, which we write as $\oeutttext{}_R$. Like \eutttext{},
\oeutttext{} is a weak simulation relation, but it additionally allows
$\oeutttext{}_R(t, \tm{trigger}\; \oom{})$ for \textit{any} interaction tree
$t$---this is the sense in which ``out-of-memory'' refines everything.

Taken altogether, these definitions lead to the following top-level, 
 definition of semantic refinement for two sets of behaviors $P$ and $Q$:
\[
  \langrefine{\vir}{Q}{P} \quad := \quad \forall t' \in Q, \exists t \in P, \hasub{t} \OR \oeutttext_{\approx_{env_{\vir}}}(t,t')
\]
Once again, we can define refinement for \vir programs $p$ and $q$ by
taking $P = \sem{p}_{\vir} \;g\;l\;sid\;m$ and $Q = \sem{q}_{\vir} \;g\;l\;sid\;m$.
\subsubsection{Refinement Theorems for $\vir^{X}$}

As mentioned above, the \vir development proves that refinement at lower levels
in the interpretation stack of Figure~\ref{fig:interp} imply refinement at later
levels (these are, intuitively, easy to prove because the less interpretation
that has been done, the \textit{stronger} the notion of refinement is).  That means we can prove:

\begin{theorem}[Level refinement]
For interpretations levels $\ell \leq \ell'$ and for any behaviors $P$ and $Q$, if \langrefine{\ell}{Q}{P} then \langrefine{\ell'}{Q}{P}.  In particular, for any $\ell$, we have \langrefine{\ell}{Q}{P} implies \langrefine{\vir}{Q}{P}.
\end{theorem}

\noindent Equally important is the ability to serially compose program refinements
\textit{within} a level of interpretation---as is needed to prove a pipeline of
program optimizations correct. To this end, we prove:

\begin{theorem}[Transitivity of refinement]
At every level $\ell$, if \langrefine{\ell}{Q}{P} and \langrefine{\ell}{R}{Q} then it is also the case that \langrefine{\ell}{R}{P}.
\end{theorem}  

\subsection{Lowering \INF{\vir} to \FIN{\vir}}
\label{sec:lowering}

The main idea in this paper is to separate compilation into two
distinct phases---there is an explicit transition from a source
language such as \INF{\vir}, with semantics using infinite memory, to
a ``target'' language such as \FIN{\vir}, which uses a finite
memory. Intuitively, when we convert an infinite program to a finite
program the \emph{only} difference in their behavior should be that
the finite program can halt with an out-of-memory event at any point,
instead of continuing execution. The semantics of \FIN{\vir} is more
constrained than that of \INF{\vir} because the finite address size
and \intptr size means that programs which allocate too much memory or
compute addresses outside of the bounds of the finite memory cannot
continue execution and must halt and trigger \OOM instead.

We can express this connection as (yet another!)  refinement. This
relation is defined in terms of $\orutttext{}_R$ (similar to how
$\oeutttext{}_R$ is used to define the single-language refinements in
Section~\ref{sec:vir-refinement}). $\orutttext{}_R$ is a heterogeneous
version of $\oeutttext{}_R$, based on the $\rutttext{}_R$ relation
between \itreesn with different event structures instead of
$\eutttext{}_R$, which operates on \itreesn with the same event
types. This is necessary as \INF{\vir} and \FIN{\vir} have events
which are parameterized by the types of addresses and \intptr values.

The
correspondence between memory configurations is given the (overloaded)
$\refineIF$ relation shown in
Figure~\ref{fig:inf-to-fin-refinement}. To express the relationship
between \INF{\vir} under-defined values and \FIN{\vir} ones, we also
need to instantiate the \tm{lift\_sbyte} function required in that
Figure.  To do so, we simply lift the \lift{p} operation (which
injects pointers) to all of the $uv$ cases---the resulting relation is
an injection that lifts a finite $uv$ to its infinite counterpart,
$\lift{uv}$. That definition allow us
to define the $\refineIF$ relation for environments too.

Putting all the pieces together, yields the following definition:

\[
  P \refineIF_{\vir} Q :=  \qquad \forall t' \in Q, \exists t \in P.\
  \hasub{t}
  \OR
                        \orutttext_{(\refineIF_{mem} \otimes \refineIF_{env})}(t,t')
\]

This definition says that for every behavior exhibited by the finite
semantics, $Q$ (as represented by the \itreen $t'$), we can find a
corresponding behavior, $t'$ in the infinite semantics, $P$. The
\itreesn that represent the behaviors should agree with each other,
either continuing indefinitely, or until both \itreesn terminate in
lock-step (by raising an error or returning a value successfully), or
until the finite \itreen raises an out-of-memory event. Finally this
relation considers \ub{}, if any \itreen in $P$ contains UB the
relation holds.

\begin{theorem}[Infinite-to-finite Top-level Refinement]
  \label{thm:inf-to-finite-refinement}
  For every \vir program $p$,\[ \INF{\sem{p}_{\vir}} \;
    \INF{g_{init}}\;\INF{l_{init}}\;\INF{sid_{init}}\;\INF{m_{init}}
    \refineIF_{\vir} \FIN{\sem{p}_{\vir}} \;
    \FIN{g_{init}}\;\FIN{l_{init}}\;\FIN{sid_{init}}\;\FIN{m_{init}} \]
\end{theorem}

This guarantees that our translation does not add any new behaviors, and that
the finite program will behave identically to the infinite one until the
programs terminate in lock-step, or the finite program runs out of memory.
Despite the apparent simplicity, this is a very technically challenging theorem
to prove for several reasons.  First, because it quantifies over \textit{all}
programs, it touches the full semantics of both \INF{\vir} and \FIN{\vir},
which, for LLVM IR, involves dozens of arithmetic, bitwise, logic, and datatype
manipulation instructions---there are literally hundreds of cases to consider.

Second, it is asking us to prove an \textit{existential} claim. Digging into
the proof,  we end up needing a lemma roughly of the form:
\[
  \forall \INF{t} \FIN{t}, \orutttext_{R_1}{(\INF{t}, \FIN{t})} -> \forall \FIN{t_2} \sem{\FIN{t}} \ni \FIN{t_2} ->
  \exists \INF{t_2}, \sem{\INF{t}} \ni \INF{t_2} \AND \orutttext_{R_2}{(\INF{t_2}, \FIN{t_2})}
\]

That is, we need to find a \INF{\vir} tree, $\INF{t_2}$, whose behaviors agree
with the \FIN{\vir} tree $\FIN{t_2}$ except for \OOM.
Ideally we would
be able to use coinduction to walk through the
$\orutttext_{R_1}{(\INF{t}, \FIN{t})}$ relation to build $\INF{t_2}$,
because that would give us the appropriate relationships between
continuations nodes in corresponding parts of the \itreesn. Unfortunately,
existentials are \textit{inductive} in Coq, so we cannot use coinduction to
extract information from this relation until the existential is
already instantiated... which is too late! We therefore have 
to define a coinductive function that lifts the finite $\FIN{t_2}$ to
the infinite $\INF{t_2}$, and then re-derive the relationship between
them.

Finally, because the semantic interpretations on both sides are defined by
layers of monadic interpreters (as in Figure~\ref{fig:interp}), the proof itself
proceeds by establishing the connection between infinite and finite semantics at
each layer, leading to many refinement lemmas, that together imply this theorem.
(There are other technical hurdles too---the \orutttext{} relation used here and
earlier is itself a non-trivial variant of the ITrees \rutttext{} mixed
inductive-coinductive definition, which requires a significant amount of
metatheory, for instance to prove transitivity, to be useful.)

%% file: optimization-examples.tex
This section explores some important program transformations enabled by our
memory model using the code examples shown in Figure~\ref{fig:opt-examples}. We have
verified refinement relations between these blocks of code, in both the infinite
language and in the finite language (where applicable). Though we have not (yet)
verified full-blown optimization passes based on these
transformations\footnote{In general, doing that would require static analysis
  and non-trivial manipulation of \vir syntax, which, while certainly doable, is
  beyond the scope of this paper.}, the semantic reasoning used in the following
refinement proofs is representative of the key ideas needed for the general case.
A notable aspect of these examples is that the infinite memory model
allows for dead allocation removal while the finite memory model does
not.

\begin{figure}[t]
\noindent  
  \begin{minipage}[t]{1.75in}
\begin{minted}[fontsize=\footnotesize]{llvm}
define void @alloca_code() {
  %ptr = alloca i64
  ret void 
}
\end{minted}
\end{minipage}
\begin{noerr}
\begin{minipage}[t]{2in}
\begin{minted}[fontsize=\footnotesize]{llvm}
define void @ptoi_code() {
  %ptr = alloca i64
  %i = ptrtoint ptr %ptr to iptr
  ret void 
}
\end{minted}
\end{minipage}
\begin{minipage}[t]{1.5in}
\begin{minted}[fontsize=\footnotesize]{llvm}
define void @ret_code() {
  ret void 
}
\end{minted}
\end{minipage}
\end{noerr}
\vspace{-3ex}
\caption{Example code for optimizations.}
\label{fig:opt-examples}
\end{figure}

The main results, each verified in Coq, are as follows:

\medskip

\noindent\textbf{Optimization 1. Dead allocation removal} (only allowed in the infinite model):

  \[
    \forall g\;l\;sid\;m. \langrefine{\vir}{\sem{\tm{@ret\_code}}_{\vir} \;g\;l\;sid\;m}{\sem{\tm{@alloca\_code}}_{\vir} \;g\;l\;sid\;m}
  \]

  Note that the twin-allocation model and CompCertS models described in
  Section~\ref{sec:remembering-memory-models} are not able to perform
  this transformation in general, unless they can verify that the
  allocation always succeeds---otherwise, removing the allocation may
  cause the program to continue executing instead of halting. This is
  not a problem in our two-phased model because allocations in the
  infinitary semantics always succeed, so we never have to worry about
  failed allocations hiding extra behaviors of the program.

\medskip
  
\noindent\textbf{Optimization 2. Removing a \tm{ptrtoint} cast} (only allowed in infinite model):

  \[
    \forall g\;l\;sid\;m. \langrefine{\vir}{\sem{\tm{@ret\_code}}_{\vir} \;g\;l\;sid\;m}{\sem{\tm{@ptoi\_code}}_{\vir} \;g\;l\;sid\;m}
  \]

  The twin-allocation and CompCertS models would be able to remove the
  \tm{ptrtoint} cast in this example, but still would not be able to remove the \tm{alloca}
  (as in the previous example). The quasi-concrete model cannot
  justify this refinement, because casting a pointer to an
  integer impacts the layout of the concrete memory and, in a finite
  setting, that could potentially result in the program halting (and thus
  removing the cast could change the behavior of the program). Again, this is
  something that the two-phased model is able to handle gracefully, as
  pointer to integer casts are essentially no-ops. The cast could be
  removed in both the finite and infinite models, but as per the previous
  example, the allocation can only be removed in the infinite.

\medskip
  
\noindent\textbf{Optimization 3. Adding an \tm{alloca}} (allowed in both the infinite and finite model):

  \[
    \forall g\;l\;sid\;m. \langrefine{\vir}{\sem{\tm{@alloca\_code}}_{\vir} \;g\;l\;sid\;m}{\sem{\tm{@ret\_code}}_{\vir} \;g\;l\;sid\;m}
  \]

  Finally, we may wish to \textit{add} an allocation to a program (certain
  optimizations may wish to cache a result, for instance). This proves tricky
  for the approach taken by CompCertS, which maintains an invariant that memory
  usage never increases after a program transformation. Both our infinite and
  finite models allow this, however, thanks to the out-of-memory refinement
  relations we've developed.

\subsection{Bounds Checking Overhead}
\label{sec:bounds-checking}

Our two-phased memory model ensures that pointer--integer casts never have an
external effect, which allows them to be removed when performing program
transformations. One might reasonably wonder, however, about the bounds checks
on \intptr arithmetic in \FIN{\vir} and whether these would impact possible
optimizations. They \textit{do}, but we believe the impact should be fairly minimal for
the following reasons.

Firstly, nearly all optimizations should be performed under \INF{\vir}
semantics, prior to lowering the program into the finitary semantics. Under the
infinitary semantics, \intptr arithmetic is just arithmetic on $\mathbb{Z}$, and
expressions involving \intptr can be optimized in the infinite world using these
unbounded integers as a model without bounds checks. Any \intptr computations
that happen to be dead can be removed prior to lowering the program into the
finite world.

All normal optimizations can occur at the infinite level, and thus the
\emph{only} optimizations necessary to do on finite programs would involve
removing the bounds checks required to trigger \OOM that are added by the
infinite to finite translation. These bounds checks can, naturally, have a
performance impact; however, we believe that they will not be a significant
impediment to the performance of real-world programs, and, in many cases,
optimizations on finitary LLVM programs should be able to remove these bounds
checks entirely. Consider the following possible use cases for \tm{ptrtoint}
casts \intptr arithmetic, which cover many real-world use cases:

\begin{enumerate}
\item Pointers cast to integers to use as a hash. \label{itm:hashing}
\item XOR doubly-linked lists. \label{itm:xorlist}
\item Using the least-significant-bit of a pointer as a flag. \label{itm:flags}
\item Indexing into allocated blocks. \label{itm:indexing}
\end{enumerate}

For (\ref{itm:hashing}), pointers can be cast to simple integer types,
like \tm{i64}, instead. The truncation does not matter in these use
cases, as the program will not cast the value back to a pointer. This
will, however, require programmers to make a choice to cast to the
appropriate integer type.

Doubly-linked lists using xor (\ref{itm:xorlist}) are an interesting
use of pointer arithmetic, however the finite \intptr values will be
64-bit values, and performing a bitwise xor cannot yield an out of
bounds value. Similarly, bitwise operations that use unused bits in
pointers as flags (\ref{itm:flags}) cannot cause an overflow either,
so bounds checking will not be necessary for these operations.

And, of course, another important case to consider is the use of \intptr
arithmetic to index into an allocated block. However, this use case should be
covered by the LLVM IR's \tm{getelementptr} operation instead, where bounds
checks are unnecessary. If \tm{getelementptr} is used to compute an out of
bounds pointer, using that pointer to perform a memory access will cause \ub{}
in the infinite semantics anyway due to mismatched provenances.

Finally, existing programming languages like Rust can achieve a great deal of
performance, despite requiring bounds checking for array
accesses~\cite{rustbounds}. We're optimistic that 1) most situations where
\intptr arithmetic will be used will fall into these cases and not require
bounds checking, 2) in rarer circumstances, other LLVM optimizations may be able
to remove the bounds checks, and 3) for any remaining bounds checks the costs
will be minimal. We believe that the flexibility our memory model allows for
optimizations prior to the finite language level will outweigh these rare costs.

%% file: memory-model-executable.tex
A formal specification of a language should be useful, in that it allows for
validating optimizations of interest, but also faithful to existing
implementations and informal specifications.
Where usefulness is the realm of formal verification, faithfulness sends us back
to a more traditional software engineering consideration: testing.
This need for validation is well identified among contributors of formal
semantics, and has even led to the development of dedicated tools and techniques
to alleviate the pain: ad-hoc usage of big-step
semantics~\cite{DBLP:conf/esop/Chargueraud13,DBLP:conf/popl/BodinCFGMNSS14}, the
K framework~\cite{kframework}, and skeletal semantics~\cite{skeletal} all notably contribute in this
direction.

\subsection{Executable Memory Models}
\label{subsec:exec-memory-model}

The \itreen{} framework~\cite{itrees}, on which we base our work, is
extremely helpful for validating such large scale semantics as
\itreesn{} can be extracted to executable code. In our case, the
memory model presented in Section~\ref{sec:memory-model-details} is
not deterministic ---a crucial necessity to faithfully characterize
memory for LLVM. Therefore it's intrinsically non-executable, as we
implement in Coq the specification monad propositionally, representing
sets $\mathcal P(A)$ as predicates \ilc{A -> Prop}.

To facilitate testing (see below), we provide proven-correct,
executable versions of the memory model.  To lighten the induced
development burden, we maintain the implementation as monadic code as
parallel as possible to the specification, which helps, in particular,
with mirroring of changes between them.

\begin{figure}
  \begin{minipage}[l]{0.48\textwidth}
    \[
      \hspace{-0.5cm}
      \small{\begin{array}{{rcl}}
        \readbe{(\varptr : \Ptr)}{} & : & \MemExec{\SByte} \\
        \writebe{(\varptr : \Ptr)}{(\varb: \SByte)}{} & : & \MemExec{\unitn} \\
        \pushfbe{} & : & \MemExec{\unitn} \\
        \popfbe{} & : & \MemExec{\unitn}
      \end{array}}
  \]
\end{minipage}
\begin{minipage}[r]{0.48\textwidth}
  \[
    \small{\begin{array}{{rcl}}
      \allocabe{(\varbs: \tlist{\SByte})} & : & \MemExec{\Ptr} \\
      \mallocbe{(\varbs: \tlist{\SByte})} & : & \MemExec{\Ptr} \\
      \freebe{(\varptr: \Ptr)}{} & : & \MemExec{\unitn}
    \end{array}}
\]
\end{minipage}
\caption{Executable memory model: low level operations}
\label{fig:memory-interface-run}
\end{figure}

Figure~\ref{fig:memory-interface-run} describes the executable
memory model interface: it precisely mimics the specification, except that it
lives in a \emph{deterministic, executable monad}:
$\MemExec{X} \defeq \MemState \to \erruboom{\MemState \times X}$.

The implementations of each of these operations closely mirrors their
specification counterparts.  They syntactically diverge significantly
only when the specification is nondeterministic, i.e., in the
$\freshp{}$ and $\mathtt{find\_bk}$ utilities needed for \tm{alloca}
and \tm{malloc}.

On the executable side, $\freshp{}$ simply uses a trivial freshness monad, which
increments a natural number to generate fresh provenance. Our current
implementation of $\mathtt{find\_block}$ is currently quite elementary, but
sufficient for our testing purpose: it looks up the largest addresses currently
allocated, and returns the range of the required size of following addresses.
More clever allocation strategies, such as those used by actual implementations
of \tm{malloc} to reduce memory fragmentation, could be implemented if relevant:
the specification only enforces that the allocated block is contiguous, and
disjoint from any other block.

  \subsubsection{Correctness of the executable memory models}
  \label{subsec:exec-correctness}

  We prove for each memory operation that its executable
  implementation is valid with respect to its specification
  counterpart.  Since these implementations are pure Coq functions,
  validity is almost defined as point-wise set membership, ensuring
  that, for any initial state, the computed result belongs to the
  specification, or that the specification contains undefined
  behavior:

  \noindent A basic memory model computation $(s : \MemExec{X})$ is valid with respect to a
  specification $(\mathcal S : \MemPropT{X})$ if:
  \[\forall \varc,\ \isbeh{\mathcal S~\varc}{\UB{}}{} \lor \isbeh{\mathcal
      S~\varc}{s~\varc}{}\] Our development proves these soundness lemmas for
  all of the memory model primitives.

%% file: testing.tex
Section~\ref{sec:integrating-into-llvm} describes the integration of our memory
model into \vir, a formal model of LLVM IR. Figure~\ref{fig:interp} also shows
the right-hand path of interpreters, which provide an executable implementation
by specializing the \textit{concretization} operation of
Section~\ref{sec:concretization} to pick default values for each \uvundef{\tau}
(for instance \uvundef{\tm{i8}} is \tm{0}).  Let us call the resulting top-level
executable program $\tm{interpret}_{\vir}$.

Using the soundness lemmas for the memory-model base operations, it is
straightforward to show that the resulting deterministic interpretation function
is a valid refinement of the semantics:

\begin{theorem}[Interpreter is sound]
  \label{thm:interpreter-is-sound}
  For all programs $p$,
  \[\langrefine{\vir}{\{\tm{interpret_{\vir}}\; p\}\;g_{init}\;l_{init}\;sid_{init}\;m_{init}}{\sem{p}_{\vir}\;g_{init}\;l_{init}\;sid_{init}\;m_{init}}\]
\end{theorem}

That is, the (singleton set) of behaviors defined by the executable interpreter
refines the semantic specification---in other words, the interpreter is ``correct.''

\subsubsection{Testing the \vir{} semantics}

The resulting \vir{} interpreter, even with the somewhat complex
memory model that manipulates symbolic bytes, is performant enough to
be able to run real LLVM IR code.  We use it on a suite of test cases
consisting of several hundred hand-written unit tests of LLVM IR
semantic features, as well as on LLVM IR code generated by compiling
source C programs. We have also experimented with using
QuickChick~\cite{denes2014quickchick} to randomly generate LLVM IR
programs that stress-test the memory model, and we can use the ability
to generate LLVM IR to instantiate parts of the Alive2~\cite{alive2}
suite as executable tests.  In all cases, we do differential testing
of the executable \vir{} model versus \tm{llc} to look for problems on
either side.  In the process of developing the memory model for this
project, such testing was invaluable to debugging the model.  It also
highlighted some ill-specified corner cases in the LLVM IR itself, for
instance, it is unclear what the \tm{getelementptr} instruction should
do when computing addresses for structures and arrays whose data
values are smaller than 8 bits and hence ``share'' an address in
memory, and \tm{extractelement} seems to have similar problems when
vector elements are smaller than 8 bits, resulting in miscompilations.